\documentclass[twocolumn]{aastex63}  
\usepackage{comment}
\usepackage{url}

\received{May 1, 2020}
\revised{November 9, 2020}
\accepted{November 10, 2020}
\submitjournal{ApJ}

\shorttitle{Small-scale flares derived by genetic algorithm}
\shortauthors{Kawai and Imada}

\begin{document}

\title{Energy distribution of small-scale flares derived using genetic algorithm}

\correspondingauthor{Toshiki Kawai}
\email{t.kawai@isee.nagoya-u.ac.jp}

\author[0000-0003-0521-6364]{Toshiki Kawai}
\affiliation{Institute for Space-Earth Environmental Research, Nagoya University, \\
Furo-cho, Chikusa-ku, Nagoya, Japan}

\author{Shinsuke Imada}
\affiliation{Institute for Space-Earth Environmental Research, Nagoya University, \\
Furo-cho, Chikusa-ku, Nagoya, Japan}



\begin{abstract}
To understand the mechanism of coronal heating, it is crucial to derive the contribution of small-scale flares, the so-called nanoflares, to the heating up of the solar corona. 
To date, several studies have tried to derive the occurrence frequency distribution of flares as a function of energy to reveal the contribution of small-scale flares. 
However, there are no studies that derive the distribution with considering the following conditions: (1) evolution of the coronal loop plasma heated by small-scale flares, (2) loops smaller than the spatial resolution of the observed image, and (3) multiwavelength observation.
To take into account these conditions, we introduce a new method to analyze small-scale flares statistically based on a one-dimensional loop simulation and a machine learning technique, that is, genetic algorithm. 
First, we obtain six channels of SDO/AIA light curves of the active-region coronal loops. 
Second, we carry out many coronal loop simulations and obtain the SDO/AIA light curves for each simulation in a pseudo-manner. 
Third, using the genetic algorithm, we estimate the best combination of simulated light curves that reproduce the observation. 
Consequently, the observed coronal loops are heated by small-scale flares with energy flux larger than that typically required to heat up an active region intermittently. 
Moreover, we derive the occurrence frequency distribution which have various power-law indices in the range from 1 to 3, which partially supports the nanoflare heating model. 
In contrast, we find that $90\%$ of the coronal heating is done by flares that have energy larger than $10^{25}~\mathrm{erg}$. 

\end{abstract}

\keywords{Sun: corona --- Sun: flares --- hydrodynamics}


\section{Introduction} \label{sec:intro}
Understanding the mechanism of the heating up of the solar corona to a few million Kelvin or above is one of the long-standing problems in solar physics. 
To date, two primary mechanisms have been proposed to explain how the corona is heated, small-scale magnetic reconnection and wave dissipation. 
In the former model, the corona heats up due to small-scale impulsive heating events, so-called nanoflares ($E \simeq 10^{24}~\mathrm{erg}$), which are related to coronal magnetic reconnections \citep[{\it e.g.,\rm}][]{Parker1988}. 
In the wave dissipation model, the corona heats up due to Alfv\'{e}n waves propagated from the surface, which are excited due to convection \citep[{\it e.g.,\rm}][]{Antolin2010}.
In recent years, much progress has been made in regard to theories and observations for both models; however, a definitive solution to this problem is yet to be achieved \citep[{\it e.g.,\rm}][]{Klimchuk2006, Klimchuk2015}.
From the perspective of the nanoflare model, quantifying the contribution of small-scale flares in the heating up of the corona is crucial to understand their contribution to coronal heating. 

\citet{Parker1983, Parker1988} proposed that small-scale flares such as microflares and nanoflares are magnetic reconnections between the coronal magnetic fields tangled by the random motion of their foot points due to the convection of the surface. 
\citet{Cargill1994} succeeded in reproducing the emission measure observation of {\it skylab} and {\it pre-skylab} by modelling the active-region closed coronal loops as a bundle of numerous elemental loops that are randomly heated by nanoflares. 
From the spectroscopic observations, \citet{Schmelz2001} and \citet{Warren2008} reported that a coronal loop consists of plasma that has a wide range of temperatures, which suggest that coronal loop is composed of finer tubes. 
The volumetric filling factor of the loop is estimated as approximately $10\%$ by \citet{Warren2008}. 
\citet{Vekstein2000} proposed the assumption that the cross-sectional area of each elemental loop can be derived from the balance between the plasma pressure inside the loop and the outer magnetic pressure. 
Based on this assumption, \citet{Sakamoto2009} revealed that the major differences between the soft X-ray ($>2~\mathrm{MK}$) and EUV ($\simeq1~\mathrm{MK}$) nanoflare-heated coronal loops are their magnetic field strength ($40~\mathrm{G}$ for soft X-ray and $8~\mathrm{G}$ for EUV) and released energies of nanoflares ($10^{24}~\textendash~10^{25}~\mathrm{erg}$ for soft X-ray and $10^{23}~\mathrm{erg}$ for EUV). 
They also estimated the volumetric filling factors of the loops as approximately $10\%$ and $70\%$ for soft X-ray and EUV, respectively. 

A microflare was observed by the hard X-ray balloon observation in 1980 for the first time \citep{Lin1984}. 
After that, from the {\it Soft X-ray Telescope} \citep[SXT:][]{Tsuneta1991} onboard the Yohkoh satellite \citep{Ogawara1991}, it was revealed that many small explosive events occur in active regions \citep{Shimizu1992, Shimizu1994}.
To validate the possibility of the coronal heating due to small-scale flares, it is necessary to derive the occurrence frequency distribution of flares as a function of the energy. 
It is known that the occurrence frequency of the flares is distributed as a power law with the following equation:
\begin{equation}
	\frac{dN}{dE}=AE^{-\alpha}
\end{equation}
where $N$, $E$, and $\alpha$ are the event number, energy of each event, and the power law constant, respectively \citep{Hudson1991}.
From this equation, the total energy released by all detected flares $P$ can be calculated as
\begin{equation}
	P=\int_{E_{\mathrm{min}}}^{E_{\mathrm{max}}}\frac{dN}{dE}EdE=\frac{A}{-\alpha+2}\left(E_{\mathrm{max}}^{-\alpha+2}-E_{\mathrm{min}}^{-\alpha+2}\right)
\end{equation}
where, $A$ is a constant.
Therefore, small-scale flares significantly heat up the corona when $E_{\mathrm{min}}$ is small enough and $\alpha$ is greater than 2.
However, in the case of solar and stellar flares ($10^{27} \lesssim E \lesssim 10^{35}~\mathrm{erg}$), $\alpha$ is estimated as approximately $1.8$ \citep{Drake1971, Datlowe1974, Lin1984, Dennis1985, Collura1988, Shakhovskaya1989}, which does not support the nanoflare heating model. 
On the other hand, the energy distributions of smaller energy flares were analyzed using various observational equipments and methods. 
In some studies, $\alpha$ was found to be greater than 2 \citep[{\it e.g., \rm}][]{Parnell2000, Benz2002}, which suggests that the frequency distribution at smaller energies is steeper and therefore may heat the corona sufficiently, while in other studies this is not the case \citep[{\it e.g.,\rm}][]{Shimizu1995, Aschwanden2000, Tajfirouze2016, Jess2019}.

The smallest event detected by satellite observation is $10^{25}~\mathrm{erg}$ \citep{Testa2014}. 
Current satellite observations cannot detect extremely small-scale flare ($E\simeq10^{23}~\mathrm{erg}$) due to lack of sensitivity. 
On the other hand, a sounding rocket experiment, the High-resolution Coronal Imager \citep[Hi-C:][]{Kobayashi2014}, detected a nanoflare that has $E\simeq10^{23}~\mathrm{erg}$ at the foot point of a coronal loop in an active region by EUV observation \citep{Testa2013}. 
Moreover, another sounding rocket, the second flight of the Focusing Optics Solar X-ray Imager \citep[FOXSI-2:][]{Christe2016} detected the plasma of the upper 10 MK without any evident X-ray flare emissions \citep{Ishikawa2017}. 
This observation indicates that the plasma is heated by nanoflares. 

By using the 0-dimensional hydrodynamic coronal loop model EBTEL \citep[Enthalpy Based Thermal Evolution of Loops][]{Klimchuk2008, Klimchuk2015, Cargill2012} and probabilistic neural network, \citet{Tajfirouze2016} estimated the power-law index $\alpha$ and the number of elemental loops in one pixel of the EUV channel of {\it Atmospheric Imaging Assembly} \citep[AIA:][]{Lemen2012} onboard the Solar Dynamics Observatory \citep[SDO:][]{Pesnell2012}. 
They concluded that $\alpha$ is approximately 1.5 and that 1000 loops are included in one pixel of an active region. 
However, they analyzed only the case of $\alpha=1.5~\mathrm{and}~2.5$, and the range of flare heating rate was very narrow (approximately from $0.01$ to $0.5~\mathrm{erg~cm^{-3}~s^{-1}}$). 
Moreover, the location where nanoflare occurs is neglected because they used the 0-dimensional model. 
According to the model of \citet{Parker1988}, nanoflares may occur anywhere along the coronal loop. 
The plasma evolution depends on the nanoflare location even in one-dimensional coronal loop model \citep[{\it e.g., \rm}][]{Reale2008}.
Therefore, to improve on this point, at least, a one-dimensional model must be used for analysis. 
The modelling of a coronal loop has been done for a few decades \citep[{\it e.g.,\rm}][]{Priest1978}. 
Generally, it is assumed that the temperature and density are almost the same as the surrounding corona, the magnetic pressure is dominant ($\beta \ll 1$), the confined plasma is a compressible fluid, and the energy is transported only along the magnetic field lines \citep[{\it e.g.,\rm}][]{Rosner1978, Vesecky1979}. 
Therefore, the evolution of the coronal loop plasma can be described using a one-dimensional model only in case neglecting the curvature, twisting, currents, waves, and non-uniform cross-section \citep[see review by][for details]{Reale2014}.

In this paper, we derive the occurrence frequency distribution of small-scale flares as a function of energy while considering the following: 
\begin{enumerate}
\item Evolution of coronal loop plasma heated by small-scale flares in the one-dimensional model
\item Elemental loops smaller than the spatial resolution of the observed image
\item Multiwavelength observation
\end{enumerate}
As far as we know, there are few studies that derive the frequency distributions of small-scale flares with taking into account all the above conditions. 
We succeeded in carrying out such an analysis by using one of the machine learning techniques called genetic algorithm (GA), which is effective at optimizing parameter combinations. 
This paper is organized as follows: 
The active region observation, which we applied in our new method, is shown in section~\ref{sec:obs}. 
We describe the set of small-scale flare heating coronal loop simulations in section~\ref{sec:sim}. 
In section~\ref{sec:ga}, we present how we derive the flare parameter and energy distribution by using GA. 
Section~\ref{sec:result} gives the obtained results. 
Finally, we summarize this paper and discuss the validity of our method and the nanoflare heating model in section~\ref{sec:summary}.

\section{Data and Observation} \label{sec:obs}
\begin{figure}[tp]
\centering
\includegraphics[width=1\linewidth]{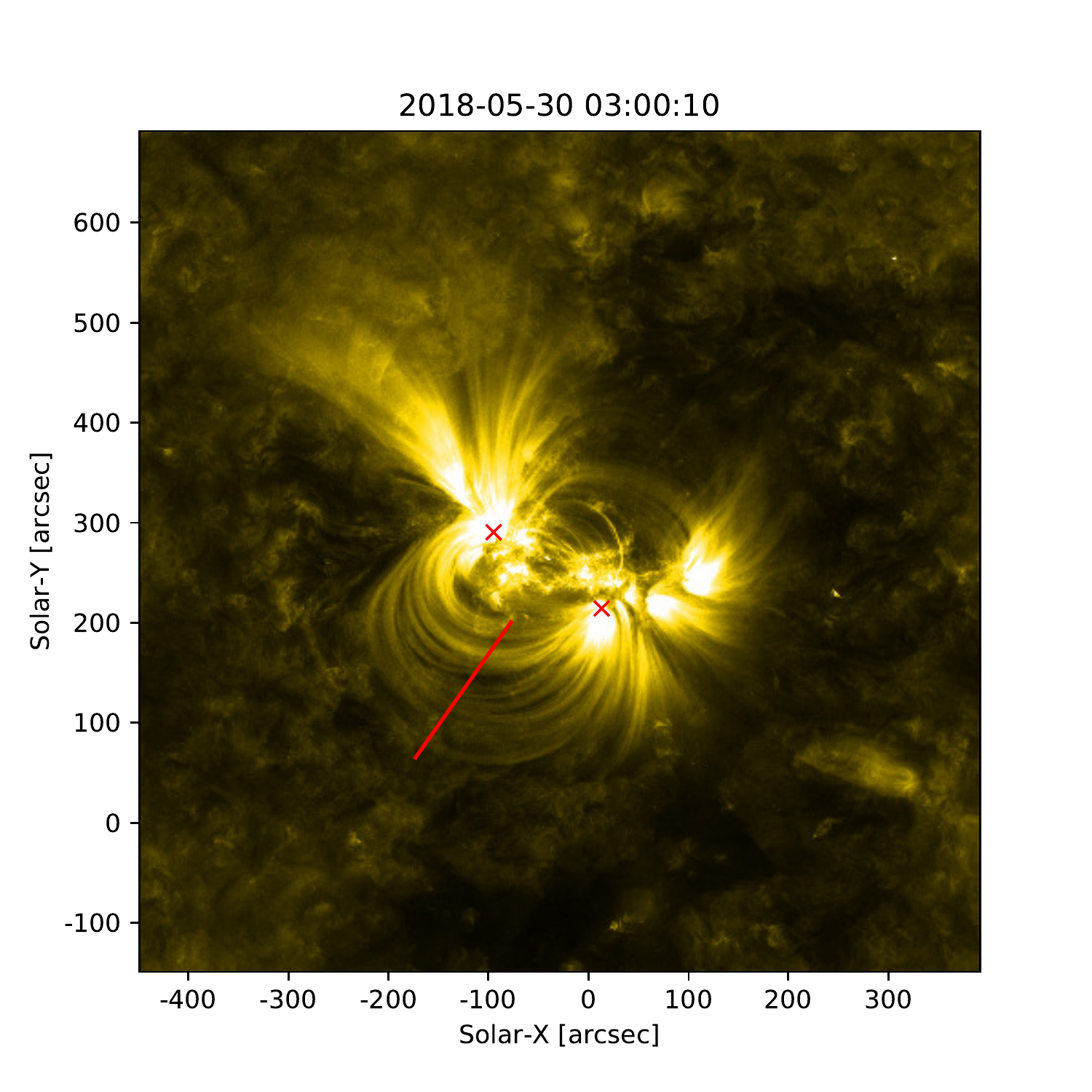}
\caption{A snapshot of NOAA active region 12712 between 03:00 -- 04:00 UT on 30 May 2018 obtained by the SDO/AIA 171 \AA. The red crosses represent the locations of loop foot points. The red line represents the vertical bisector of both foot points. \label{fig:AR}}
\end{figure}
\begin{figure}[tp]
	\centering
	\includegraphics[width=1\linewidth]{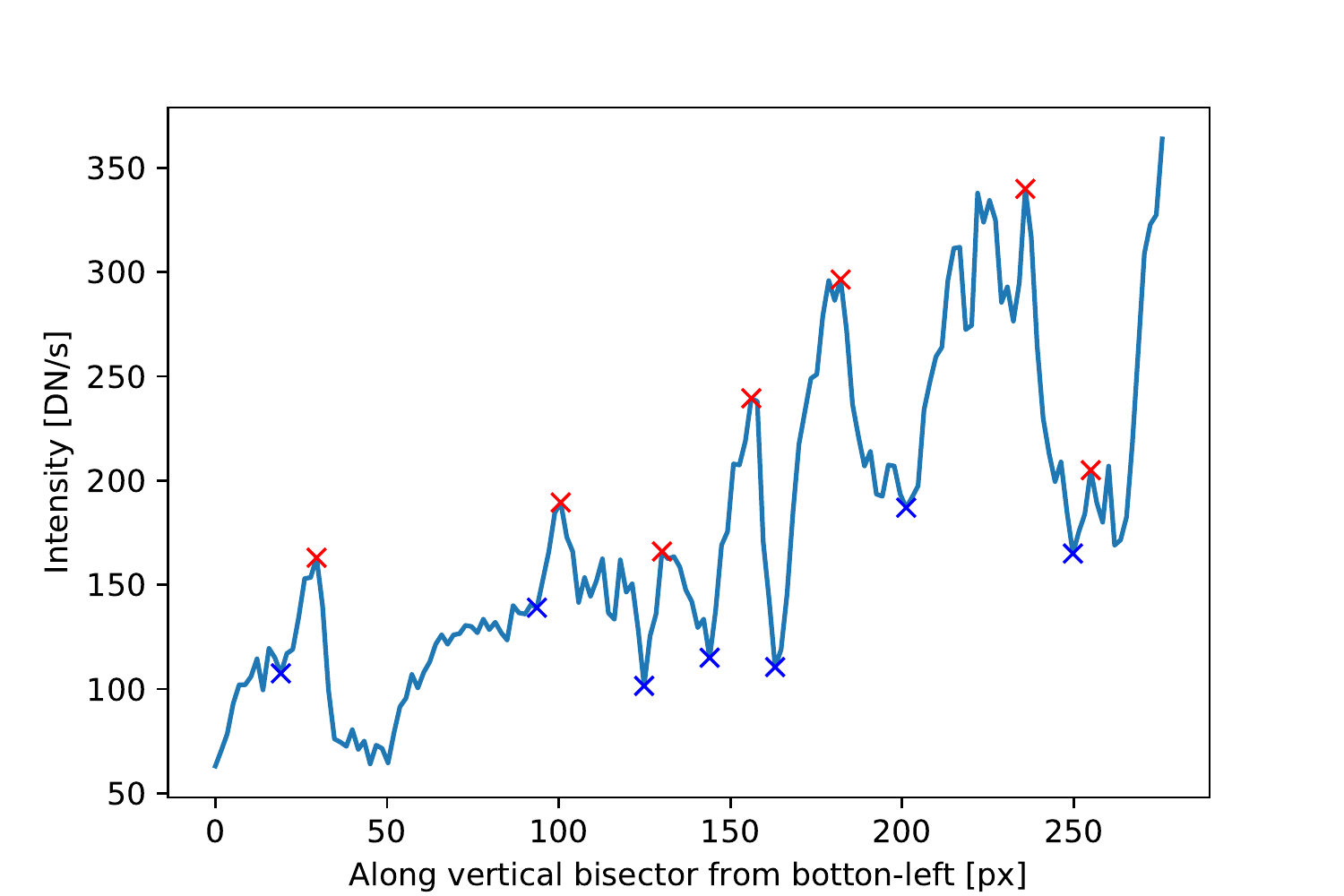}
	\caption{Intensity distribution along the vertical bisector of loop foot points (Red line in figure~\ref{fig:AR}). Red and blue crosses represent peaks and bottoms of large enhancements, respectively.}
	\label{fig:along_vb}
\end{figure}

In this study, we use the coronal EUV images of NOAA active region 12712 between 03:00 and 04:00 UT on 30 May 2018, using six filters of the SDO/AIA, which consists of 94 ($\log{T} \simeq 6.8$), 131 ($\log{T} \simeq 7.0$), 171 ($\log{T} \simeq 5.8$), 193 ($\log{T} \simeq 6.0$ for quiet region and $\log{T} \simeq 7.3$ for flaring region), 211 ($\log{T} \simeq 6.3$), and 335 ($\log{T} \simeq 6.4$)~\AA~\citep{Lemen2012}. 
This active region is suitable for small-scale flare analysis because it is relatively calm and produces no eruptions greater than a GOES B-class flare. 
The spatial and time resolution of each filter is 0.6'' and 12 s, respectively. 
The SDO/AIA data are calibrated by the \verb|aia_prep| routine in SolarSoftWare \citep[SSW:][]{Freeland1998}.
We obtain light curves at the loop tops for our analysis.
Figure~\ref{fig:AR} shows the observed active region obtained from the AIA 171 \AA~filter.
The red crosses and red line indicates the foot points of coronal loops and their vertical bisector, respectively. 
This red line corresponds to the observation locations. 
The length and location of the red line is defined to avoid the active region core and is extended to where the existence of the loop can be visually confirmed.

Figure~\ref{fig:along_vb} shows the intensity distribution along the red line in figure~\ref{fig:AR}. 
We regard spikes in this profile as coronal loops. 
Each loop we detect is enhancements above $3\sigma_\mathrm{ph}$, where $\sigma_\mathrm{ph}$ represents photon noise. 
In this study, we estimate $\sigma_\mathrm{ph}$ as the square root of mean intensity along the red line. 
The red and blue crosses in figure~\ref{fig:along_vb} represents the peak and bottom of each large spikes. 
We estimate the minimum coronal loop radius as the minimum FWHM of these which is 5.2 pixels along the vertical bisector. 
Therefore, light curves are summed in each $4 \times 4$ pixels along the red line ($5.2/\sqrt{2} \simeq 4$).

\begin{figure}[tp]
\centering
\includegraphics[width=1\linewidth]{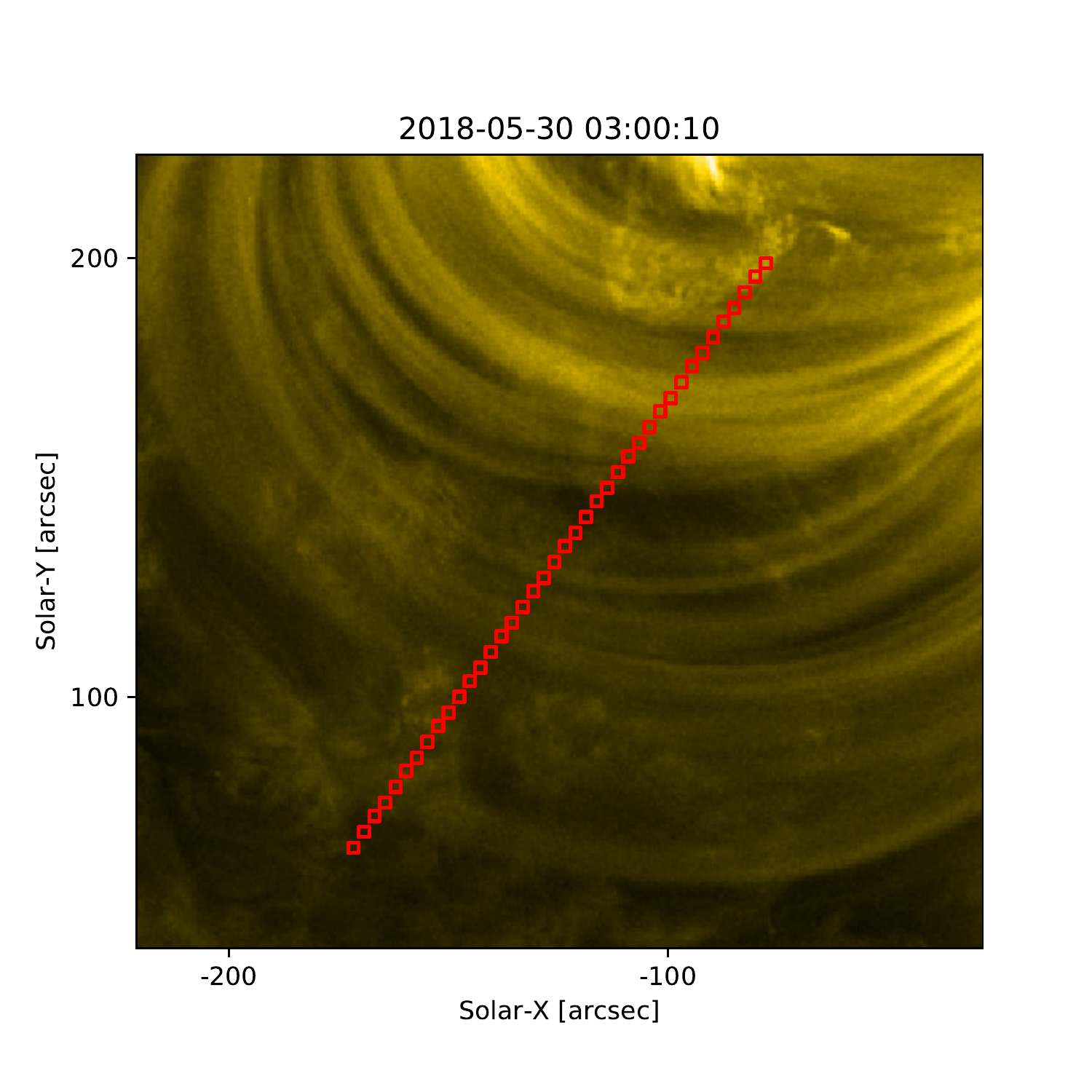}
\caption{Trimmed image of figure~\ref{fig:AR}. Forty red squares indicate the area where the light curves are obtained.}
\label{fig:obslocs}
\end{figure}

Figure~\ref{fig:obslocs} presents a cropped map of figure~\ref{fig:AR}. 
Each red square indicates pixels used in the light curve. 
The size of each square is $4 \times 4$ pixels as described above.
The number of the observation locations is 40. 

The latitude of this active region is approximately 15 deg north. 
To track the movement of this active region due to the rotation, we calculate the apparent velocity as follows:
\begin{equation}
	v_\mathrm{app} = \frac{2 \pi R_\odot \cos{\phi}}{360}v_\mathrm{rot} ~\mathrm{km~day^{-1}}
\end{equation}
where, $R_\odot$, $\phi$, and $v_\mathrm{rot}$ represent the solar radius, latitude, and the rotation rate. 
According to the result of \citet{Snodgrass1990}, the rotation rate of this active region is estimated as approximately $14.54~\mathrm{deg~day^{-1}}$. 
Therefore, $v_\mathrm{app}$ can be estimated as $1.7 \times 10^5~\mathrm{km~day^{-1}}$
We assume that 1 arcsec along the solar-X at this active region is equivalent to 730 km. 
This assumption is reasonable because the longitude of this active region is almost zero. 
We calculate the apparent velocity of this active region as $9.8~\mathrm{arcsec~h^{-1}}$. 
The apparent motion is corrected for by moving the pixels in the light curve accordingly.

\begin{figure*}[tp]
	\centering
	\includegraphics[width=1\linewidth]{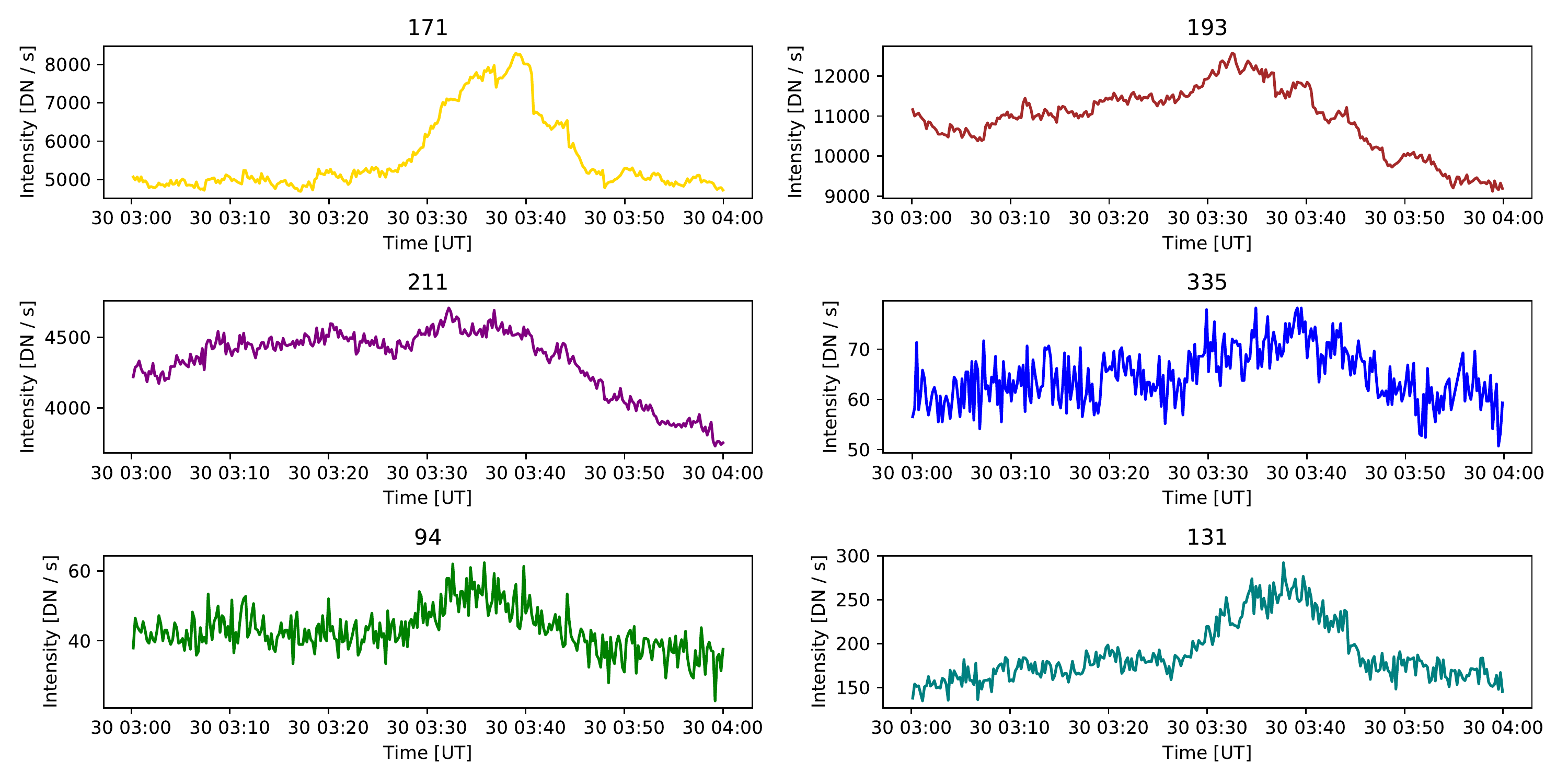}
	\caption{An example of light curves of one observational region (top-right red square in Figure~\ref{fig:obslocs}) obtained from six filters of SDO/AIA.}
	\label{fig:sample_lc}
\end{figure*}

We obtain the light curves of the six filters in each red box. 
For example, Figure~\ref{fig:sample_lc} represents the observed light curves in one red box. 
Some enhancements can be seen in the cooler channels (171, 193, 211 \AA), while the intensities of hotter channels (335, 94, 131 \AA) are very low and noisy. 
The loop lengths can be roughly estimated as $200~\textendash~400~\mathrm{Mm}$. 


\section{Numerical Simulation} \label{sec:sim}
We calculate the evolution of a coronal loop plasma heated by flares with a one-dimensional model. 
In this study, we use CANS (coordinated astronomical numerical software \footnote[1]{Details are available at the website \url{http://www.astro.phys.s.chiba-u.ac.jp/netlab/astro/index2-e.html}. We used the version of 21 September 2019 distributed at \url{http://www-space.pdf.s.u-tokyo.ac.jp/~yokoyama/etc/cans/index-e.html}.}) 1D solar flare package. 
The fundamental equations are as follows: 
\begin{eqnarray}
\frac{\partial}{\partial t}(\rho S) + \frac{\partial}{\partial x}\left(\rho V_x S\right) = 0 \\ 
\frac{\partial}{\partial t}\left(\rho V_x S\right) + \frac{\partial}{\partial x}\left[\left(\rho V_x^2 + p\right)S\right] = \rho g S \\ 
\nonumber
\frac{\partial}{\partial t}\left[\left(\frac{p}{\gamma - 1} + \frac{1}{2}\rho V_x^2\right)S\right] + \\
\nonumber
\frac{\partial}{\partial x}\left[\left(\frac{\gamma}{\gamma - 1}p + \frac{1}{2}\rho V_x^2\right)V_x S - \kappa \frac{\partial T}{\partial x}S\right]= \\
\left(\rho g V_x + H - R + H_f\right)S \\ 
p = \frac{k_B}{m}\rho T
\end{eqnarray}
where, $p$, $T$, $v_x$, $\rho$, $\gamma=5/3$, $S$, $g$, $H$, $R$, $H_f$, $k_B$, $\kappa$, and $m$ represent the pressure, temperature, plasma velocity along the loop, density, heat capacity ratio, cross-sectional area, gravitational acceleration, static heating, radiative cooling, flare heating, Boltzmann constant, thermal conductivity, and mean particle mass, respectively. 
The simulation assumes that the length and cross section of the loop do not change with time; the cross section is uniform along the loop; the flow is inviscid and compressible; and the location where the flare occurs is fixed at the loop top. 
The Spitzer thermal conductivity \citep{Spitzer1956} and gravity are taken into account as follows. 
\begin{eqnarray}
\kappa = \kappa_0 T^{5/2}\\
g=g_0\cos\left(\frac{\pi}{2L}x\right)
\end{eqnarray}
where, $\kappa_0=10^{-6}~\mathrm{cgs}$, $g_0=270~\mathrm{m/s^2}$ and $L$ represent thermal conductivity strength, a gravitational acceleration at the photosphere, and the half loop length. 

An approximation to correct for the effects of high-density plasma is included in the radiative cooling model as following equations: 
\begin{eqnarray}
R=\rho^2\Lambda_\rho(\rho)\Lambda(T)\\
\Lambda_\rho(\rho)=\frac{\rho_{\mathrm{cl}}}{\rho}\tanh\left(\frac{\rho}{\rho_{\mathrm{cl}}}\right)
\end{eqnarray}
where,  $\rho_{\mathrm{cl}}=10^{12}~\mathrm{cm^{-3}}$.
The radiative cooling function $\Lambda(T)$ is approximated by an algebraic function as follow:
\begin{eqnarray}
\Lambda(T) \approx \Lambda_0 10^{\Theta(T)}\\
\theta \equiv \log_{10}\left( \frac{T}{T_{\mathrm{cl}}}\right)\\
\Theta(\theta) = 0.4\theta - 3+3\times \frac{2}{e^{1.5(\theta + 0.08)}+e^{-2(\theta+0.08)}} 
\end{eqnarray}
where, $\Lambda_0=8 \times 10^{-22}~\mathrm{cgs}$ and $T_{\mathrm{cl}}=2 \times 10^{5}~\mathrm{K}$. 

This simulation includes not only the corona but also the transition region and chromosphere. 
The flare energy input is represented by the following equations:
\begin{eqnarray}
H_f = H_{f0} \cdot q(t) \cdot f(x) \cdot g_1(x)\cdot g_2(x)\\
q(t) = \frac{1}{4}\left\{ 1 + \tanh{\frac{t-t_{i0}}{0.1 \tau_0}}\right\}\left\{1-\tanh{\frac{t-t_{i1}}{0.1 \tau_0}}\right\}\\
f(x) = \frac{1}{\sqrt{2 \pi}} \exp{\left[ -\frac{(x-x_i)^2}{2w_f^2}\right]} \\
g_1(x) = \frac{1}{2} \left\{ 1 + \tanh{ \frac{x-20 \mathcal{H}_0}{3 \mathcal{H}_0}} \right\} \\
g_2(x) = \frac{1}{2} \left\{ 1 - \tanh{\frac{x-(2L-20\mathcal{H}_0)}{3 \mathcal{H}_0}} \right\}
\end{eqnarray}
where, $w_f=6000~\mathrm{km}$, $\mathcal{H}_0=200~\mathrm{km}$, and $\tau_0=20~\mathrm{s}$ represent the width along the loop of the flare, the scale height, and sound wave traveling time at the surface ($x=0$), respectively. 
$t_{i0}$, $t_{i1}$, and $x_i$ represent the beginning and finishing time and location of heating by the $i$th flare, respectively.
$q(t)$ is a function of time to make the heating impulsive. 
The role of $g_1(x)$ and $g_2(x)$ is to prevent the heat pulse from entering directly into the chromosphere.
This simulation uses the modified Lax--Wendroff scheme which is second-order accurate in both space and time. 
The boundary conditions at $x=0$ and $x=2L$ are as follows:
\begin{eqnarray}
\frac{\partial \rho}{\partial x} = 0\\
\frac{\partial p}{\partial x} = 0\\
V_x=0
\end{eqnarray}
This simulation setup is mostly the same as that of \citet{Hori1997, Imada2012, Kawai2020}. 

\begin{table}[tpb!]
	\centering
	\caption{Free parameters in the simulation and their ranges}	
	  \begin{tabular}{rrr}
	  Parameter & Minimum & Maximum \\ \hline
	  Heating rate~$[\mathrm{erg/s/cm^3}]$ & $10^{-2}$ & $10^2$  \\
	  Flare duration~[$\mathrm{s}$] & 1 & 300 \\
	  Flare occurrence time~[$\mathrm{s}$] & $t=0$ & $t=5600$ \\
	  Flare location & $x=0$ & $x=2L$ \\
	  Loop length~[$\mathrm{Mm}$] & 100 & 500 \\
	   \end{tabular}
	\label{tab:param}
\end{table}

\begin{figure*} [tp]
\centerline{\includegraphics[width=0.8\textwidth]{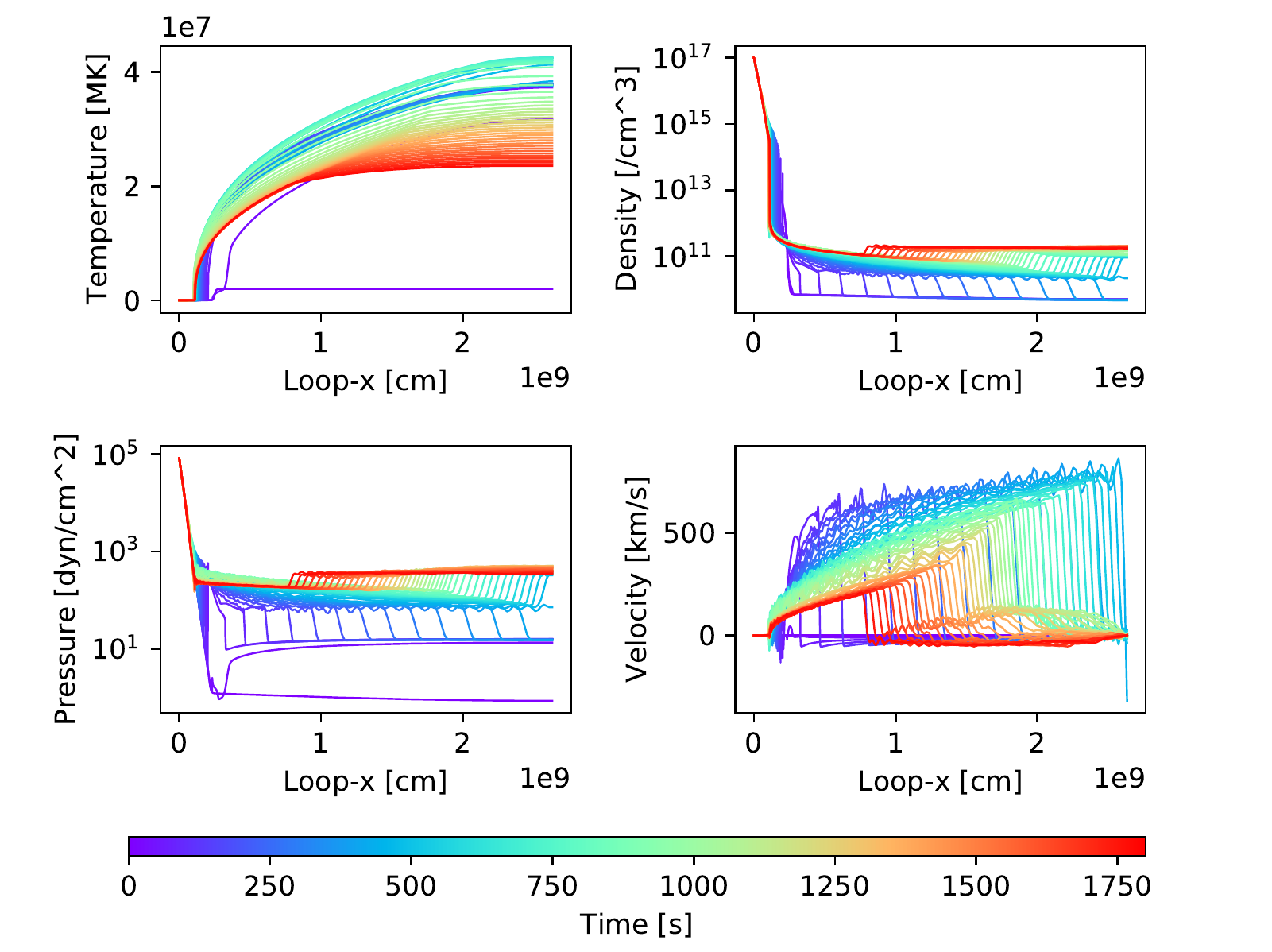}}
\caption{Results of the hydrodynamic simulation. Each panel presents the time evolution of the temperature distribution (left-top), density (right-top), pressure (left-bottom), and plasma velocity along the loop (right-bottom) with 30-s cadence. Each horizontal axis represents the loop coordinates from the surface to the loop top. Line color indicates the progress of time (from blue to red).}\label{fig:cans}
\end{figure*}

To demonstrate the simulation, we show a simple example in figure~\ref{fig:cans}. 
Each panel presents the temporal variation and spatial distribution of temperature, density, pressure, and plasma velocity along the loop, respectively. 
As each horizontal axis indicates coordinates along the loop, the left-hand edge ($x=0$) is the surface and the right-hand edge is the loop top. 
The region where the temperature and pressure change rapidly ($x\simeq 0.3$ Mm) is the transition region, and the chromosphere is in the left of this region.
Line color indicates the progress of time in the simulation, from blue to red. 
This result is for a single flare occurrence at the loop top at the beginning of the simulation. 
When a flare occurs, the temperature of the loop top is increased and is transported to the foot points of the loop by thermal conduction. 
Then, the temperature and pressure of the chromosphere are rapidly increased by the incoming high-temperature plasma. 
As a result, high-density plasma in the chromosphere is ejected into the corona by the pressure gradient force; this is referred to as chromospheric evaporation.
Consequently, the coronal loop is filled with high-density plasma, and emits soft X-ray and EUV irradiance. 

To use various results in GA later, we carry out the simulation with random flare heating rates, heating duration, occurrence time, occurrence location, number, and loop length. 
Table~\ref{tab:param} represents the parameters and their possible ranges. 
According to the study of \citet{Testa2014}, an energy flux of a nanoflare ($10^{24} \lesssim E \lesssim 10^{25}~\mathrm{erg}$) is approximately $7 \times 10^7$ -- $2 \times 10^9~\mathrm{erg/s/cm^2}$.
Therefore, the range of flare heating rate in the simulation is roughly $10^{-0.9}$ -- $10^{0.5}~\mathrm{erg/s/cm^3}$ because the width along the loop of flare is $6000~\mathrm{km}$. 
However, to include weaker and stronger heating, we defined the range of heating rate as $10^{-2}$ -- $10^{2}~\mathrm{erg/s/cm^3}$.
The heating duration of a nanoflare in \citet{Testa2014} is from 10 to 30 s.
We define the range of heating duration in our simulation from 1 (10 times shorter than the minimum) to 300 (10 times longer than the maximum) s. 
\citet{Testa2014} also suggested that the duration of heating event is similar to that of transition region brightenings. 
The heating duration range in our simulation almost covers the typical lifetimes of compact transition region brightenings, that is from a few tens of seconds to several minutes \citep{Tiwari2019}.
The probability distributions of heating rate and flare duration are uniform in a logarithmic scale while those of flare occurrence time, location, and loop length are uniform in their ranges. 
This maximum number of flares is roughly determined to distinguish each heating event in each run even when flares have the longest duration.
The wider the parameter range, the more various the simulation results, but the more calculations are required to reproduce the observed light curves well. 
These parameter ranges are wider than those of \citet{Tajfirouze2016} and enough to reproduce the observations because we focus on relatively calm coronal loops.
The cadence of the simulation output is 2 s. 

\begin{figure*}[tp]
	\centering
	\includegraphics[width=0.8\linewidth]{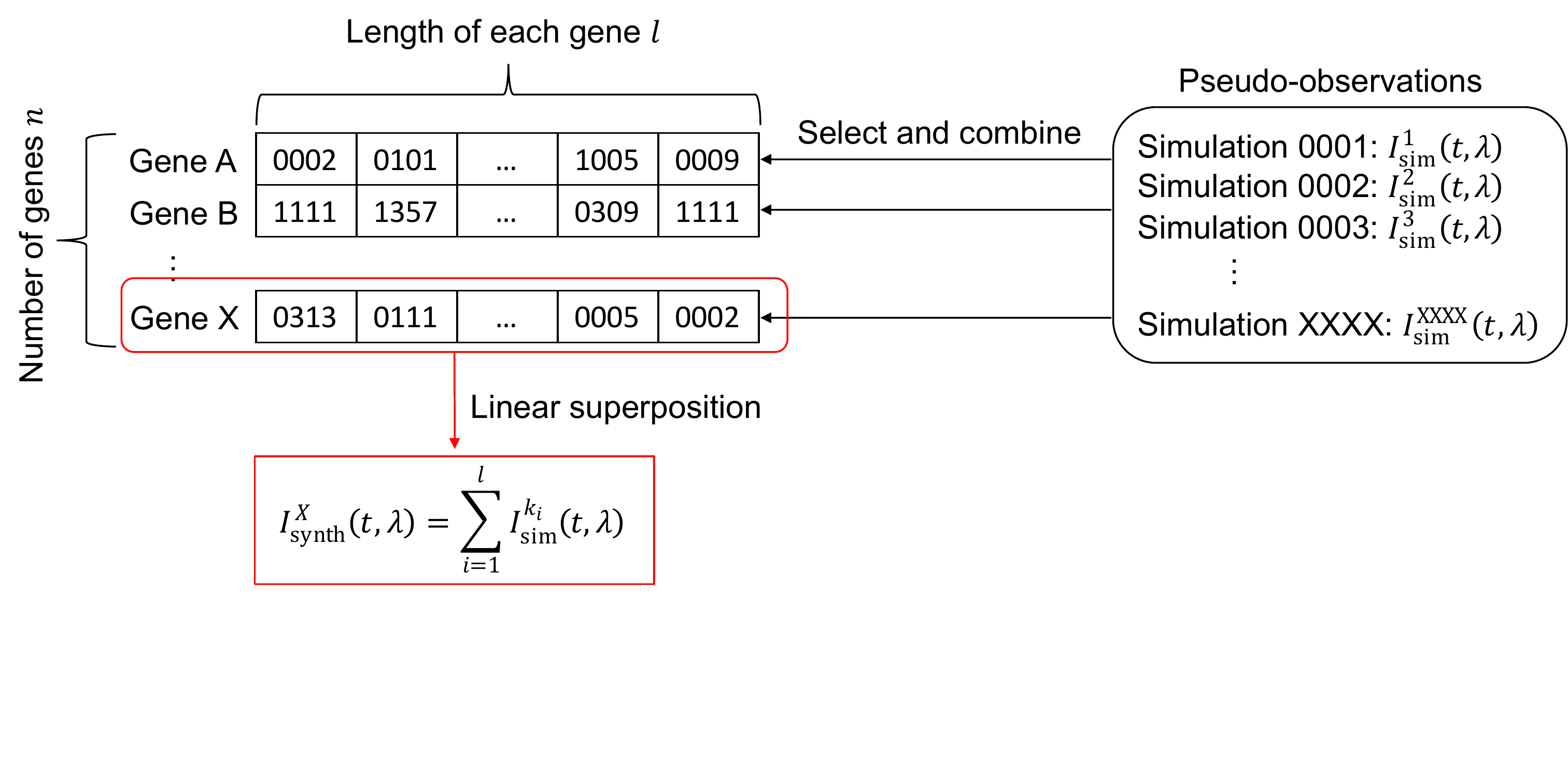}
	\vspace{-15mm} 
	\caption{A schematic architecture of a gene used in this paper. Each gene represents a synthetic light curve $I_\mathrm{synth}(t, \lambda)$, which is described as a linear superposition of $l$ of simulated light curves.}
	\label{fig:ga}
\end{figure*}

The original CANS solar flare package calculates only half of the loop (foot point to top) under the symmetry assumption as shown in figure~\ref{fig:cans}. 
his assumption should work well when a flare occurs at the top like in figure~\ref{fig:cans}. 
However, according to the Parker's model, nanoflares can occur everywhere along the loop and this symmetric assumption might be violated.
Therefore, in this study, we extend it to calculate the whole loop (one foot point to another foot point) to consider various heating locations. 
Moreover, to simplify the discussion, we remove static heating ($H=0$) and assume the loop is heated by only flares. 
There are two reasons for this assumption. 
One is that the mechanism and intensity of background heating is also unclear, therefore, another assumption is needed anyway. 
The second reason is that the contribution of background heating must be relatively small even when the loop is heated by small-scale flares.

In some conditions, the calculation is broken due to an overheating or overcooling of the loop. 
We do not use simulation results, which has density of over $10^{12}~\mathrm{cm^{-3}}$ around the loop top. 
This high density loop is sometimes caused by overcooling due to the imbalance between the flare heating and radiative and conductive cooling. 
This imbalance mostly occurs when the flare heating is too large.

Because the observation duration is one hour, as described in Section~\ref{sec:obs}, the simulation also needs to be calculated for one hour. 
However, to make various initial conditions, we calculate time in advance of observation duration. 
As can be seen in the simulation results, the variety of light curve becomes the maximum around $t=2000~\mathrm{s}$ even for the longest loops. 
Therefore, we calculate for 5600 s and remove the first 2000 s for the optimization.
As describe above, the number of flares are 15 in each run, however, some or all of them can occur during the first 2000 s. 
Therefore, the range of number of flares is from 0 to 15 during the latter one hour calculation. 
In this study, we ran approximately 5,000 random simulations. 

As described in section~\ref{sec:obs}, we obtain a light curve of the SDO/AIA $4 \times 4$ pixels ($\simeq 1.8 \times 1.8~\mathrm{Mm^2}$) at the loop top by the observation. 
Therefore, we derive the SDO/AIA light curve in the $1.8~\mathrm{Mm}$ range at the loop top of each calculation as well. 
Using the temperature response function of each SDO/AIA channel \citep{Lemen2012}, we obtain the intensities in each time and grid from the simulation result. 
The response functions are available from the SolarSoftWare \citep[SSW:][]{Freeland1998} procedure \verb|aia_get_response.pro|.
We assume the emission measure to be $\rho^2$ and derive the light curve per cross-sectional area ($\mathrm{DN~s^{-1}~cm^{-2}}$) for each calculation. 
We assign the identification to each run to use them for optimization, which is described in the next section. 

\section{Genetic Algorithm} \label{sec:ga}
We estimate the best combinations of the results of simulations that reproduce actually observed coronal loop light curves by using GA. 
GA is a machine learning technique that is effective at solving optimization and searching problems and is based on the concept of Darwin's theory of evolution. 
``Genes'' and an ``environment'' are necessary for GA.
GA can search the gene that is optimized best for the environment through natural selection, crossover, and mutation. 
In this study, each gene is a list of the simulation IDs that are assigned in section~\ref{sec:sim}. 
Each gene expresses the synthetic light curve by linear superposition of its simulation results. 
Therefore, a synthetic light curve $I_\mathrm{synth}^X$ of a gene $X$ can be described as follows:
\begin{equation}
	I_\mathrm{synth}^X(t, \lambda) = \sum_i^l I_\mathrm{sim}^{k_i}(t, \lambda)
\end{equation}
where, $l$ and $I_\mathrm{sim}^{k}$ represent the length of gene (the number of simulated loops to combine) and simulated light curve of simulation ID $k$, 
Figure~\ref{fig:ga} represents the schematic organization of the genes. 
In this paper, gene length $l$ and the number of gene $n$ are 100 and $500$, respectively.
$l$ is defined based on the analysis of \citet{Tajfirouze2016}, which suggests that the number of loops in each SDO/AIA pixel is approximately 1000. 
However, only one flare occurs on each coronal loop in their study while at most 15 flares occur on each loop in our simulation. 
Therefore, we roughly estimate the number of elemental loops in each red square in figure~\ref{fig:obslocs} as 100.
The accuracy of reproduction of light curves increases as the number of genes $n$ increases. 
We define $n=500$ based on available computational resources. 
However, even if $n$ is smaller, the reproducibility can be improved by increasing the number of iteration. 

\begin{figure}[tp]
	\centering
	\includegraphics[width=1\linewidth]{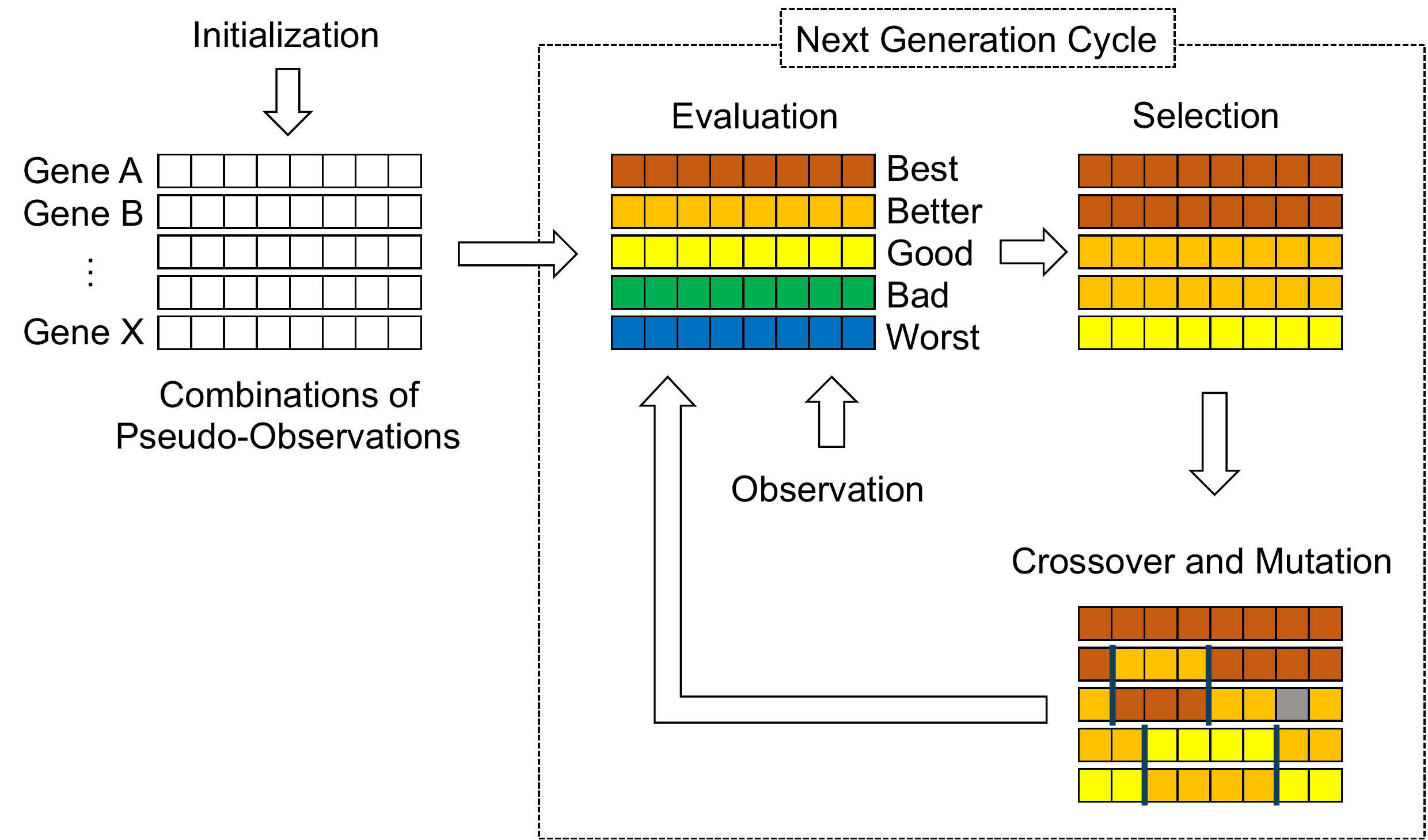}
	\caption{A schematic flow of GA. Firstly, initial genes generated randomly are evaluated based on the correlations between observed $I_\mathrm{obs}(t, \lambda)$ and synthetic $I_\mathrm{synth}(t, \lambda)$ light curves. Second, based on the evaluation, better genes are left or duplicated probabilistically, while worse genes are eliminated. Third, the remaining genes swap their information partially with each other. Sometimes a mutation occurs to avoid a local minimum. The best gene that reproduces the observed light curves can be estimated by continuing this iteration.}
	\label{fig:iter}
\end{figure}

The main components of GA are a natural selection, crossover, and mutation. 
Figure~\ref{fig:iter} schematically shows the procedures. 
Genes have a random list of IDs at the beginning. 
First, in natural selection, while better genes are left or multiplied, worse genes are eliminated stochastically. 
Each gene is evaluated by the averaged correlation between observed and synthetic light curves in six AIA filters. 
Therefore, the selection probability of gene $i$ can be described as follows:
\begin{equation}
	p_i = \frac{c_i}{\sum_{k=1}^{n} c_k}
\end{equation}
where, $c_i$ represents the averaged correlation of gene $i$. 
Genes in the next generation are decided by this probability. 
However, in this case, even the best gene among the generation has a possibility to be eliminated. 
Therefore, we leave at least one of the genes which have the highest correlation to the next generation, which is typical for GA. 
This can accelerate the learning while it may increase the likelihood of getting a local solution due to the lack of variety of genes.

Second, in crossover, we search for new better solutions by swapping information partially between the pair of the genes. 
In this study, we choose two-point crossover for this procedure.
We randomly choose pairs of genes and two crossover points (black vertical lines in figure~\ref{fig:iter}) from within the genes. 
The IDs between the crossover points are exchanged with each other. 
These new genes are populations for the next generation. 
$90~\%$ of all genes are crossed over in each generation. 

Third, some genes are mutated to maintain diversity and avoid local minima. 
In a mutation, a few IDs in the genes are randomly replaced to the another ID with a low probability. 
In this paper, $1~\%$ of all IDs in the genes are mutated in each generation. 

We estimate the gene that can best reproduce the observed light curves by iterating these procedures 1,000 times for each observational area. 
The gene is a list of simulation results and we know the inputs of flare parameters of each simulation. 
Therefore, we can estimate the combination of flare parameters that reproduce observational results best by GA. 

When synthetic light curves $I_\mathrm{synth}$ are satisfactorily correlated to observational results $I_\mathrm{obs}$, the relationship between them can be written as follows:
\begin{equation} \label{eq:A}
	 I_\mathrm{obs}(t, \lambda) \simeq A(\lambda) \times I_\mathrm{synth}(t, \lambda) + B(\lambda)
\end{equation}
where, $A$ and $B$ are coefficients derived by least squares. 
$A$ and $B$ indicate the ratio of amplitudes of fluctuations between $I_\mathrm{obs}$ and $I_\mathrm{synth}$ and the background component of $I_\mathrm{obs}$, respectively. 
As mentioned above, $I_\mathrm{sim}$ is the time series of emissions from the coronal loop whose cross-sectional area is $1~\mathrm{cm^2}$. 
Therefore, the cross-sectional area of elemental coronal loops in each observation region can be estimated as $\Delta S_\mathrm{loop}(\lambda)=A(\lambda)~\mathrm{[cm^2]}$. 

\section{Result} \label{sec:result}
\begin{figure}[tp]
	\centering
	\includegraphics[width=1\linewidth]{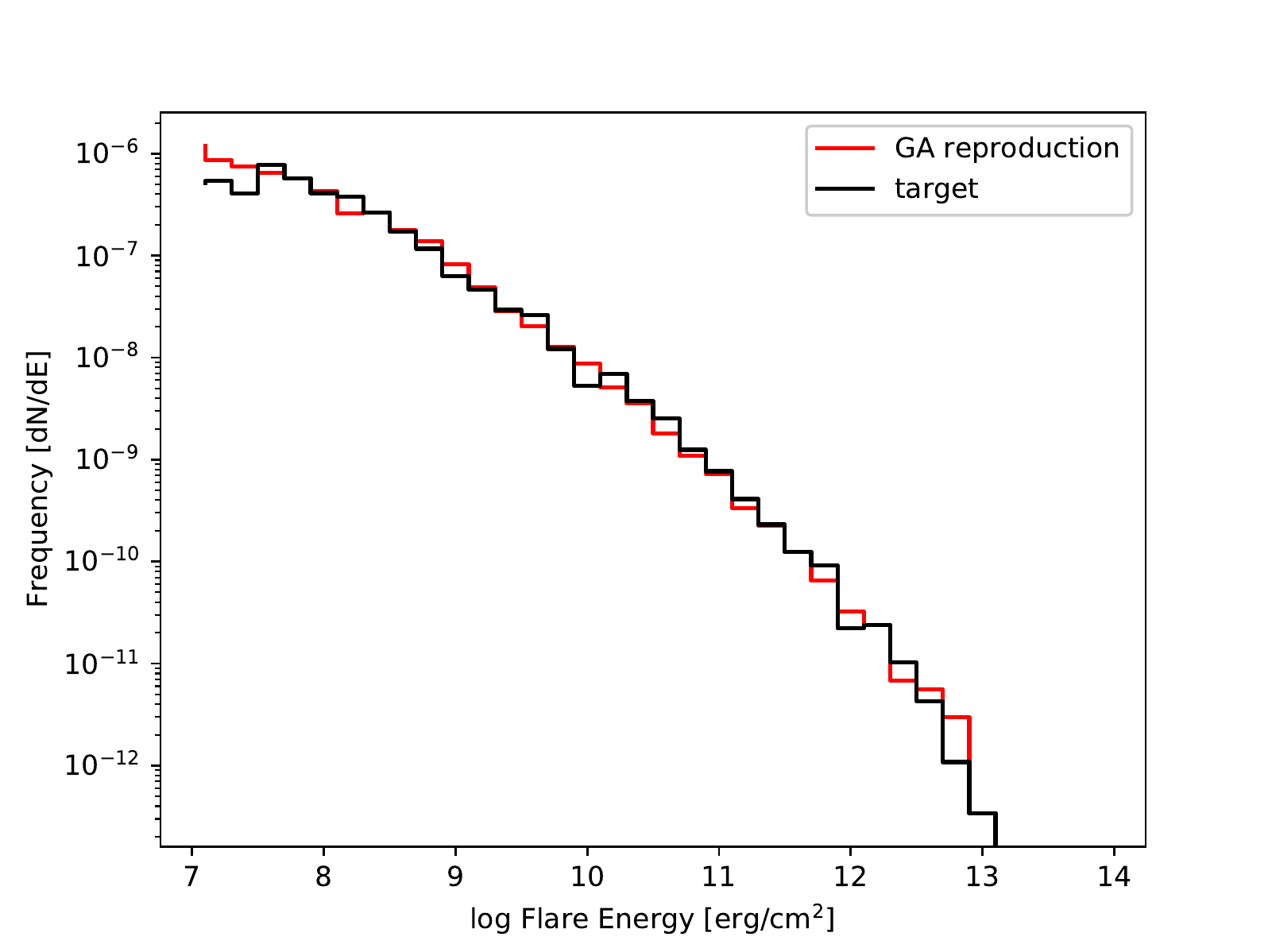}
	\caption{An example of frequency distribution of flares as a function of energy. Black and red lines indicate the distribution of input of the target simulations and of estimation by the GA, respectively.}
	\label{fig:testdist}
\end{figure}

At first, to verify the robustness of our procedure, we test whether the GA can estimate the correct combination of simulations from light curves generated by random set of genes. 
We create SDO/AIA synthetic light curves from 100 sets of simulation results as a target. 
The created light curves have a random gaussian noise whose standard deviation is $10 \%$ of the intensity. 
The estimated genes are not exactly the same with the target gene, however, the derived distribution of flares are very similar to the input of that. 
Figure~\ref{fig:testdist} represents an example of occurrence frequency distribution of flares in unit of cross sectional area. 
Red and black histograms represent distributions of GA reproduction and input of target simulations. 
In this case, the mean correlation between target and reproduced light curves is approximately 0.97.
We continue this test 25 times to define the uncertainty of this method. 
We define the uncertainty of the occurrence frequency distribution, $\sigma_f$, as a standard deviation of the difference between target and reproduced one in each energy bin. 

\begin{figure}[tp]
	\centering
	\includegraphics[width=1\linewidth]{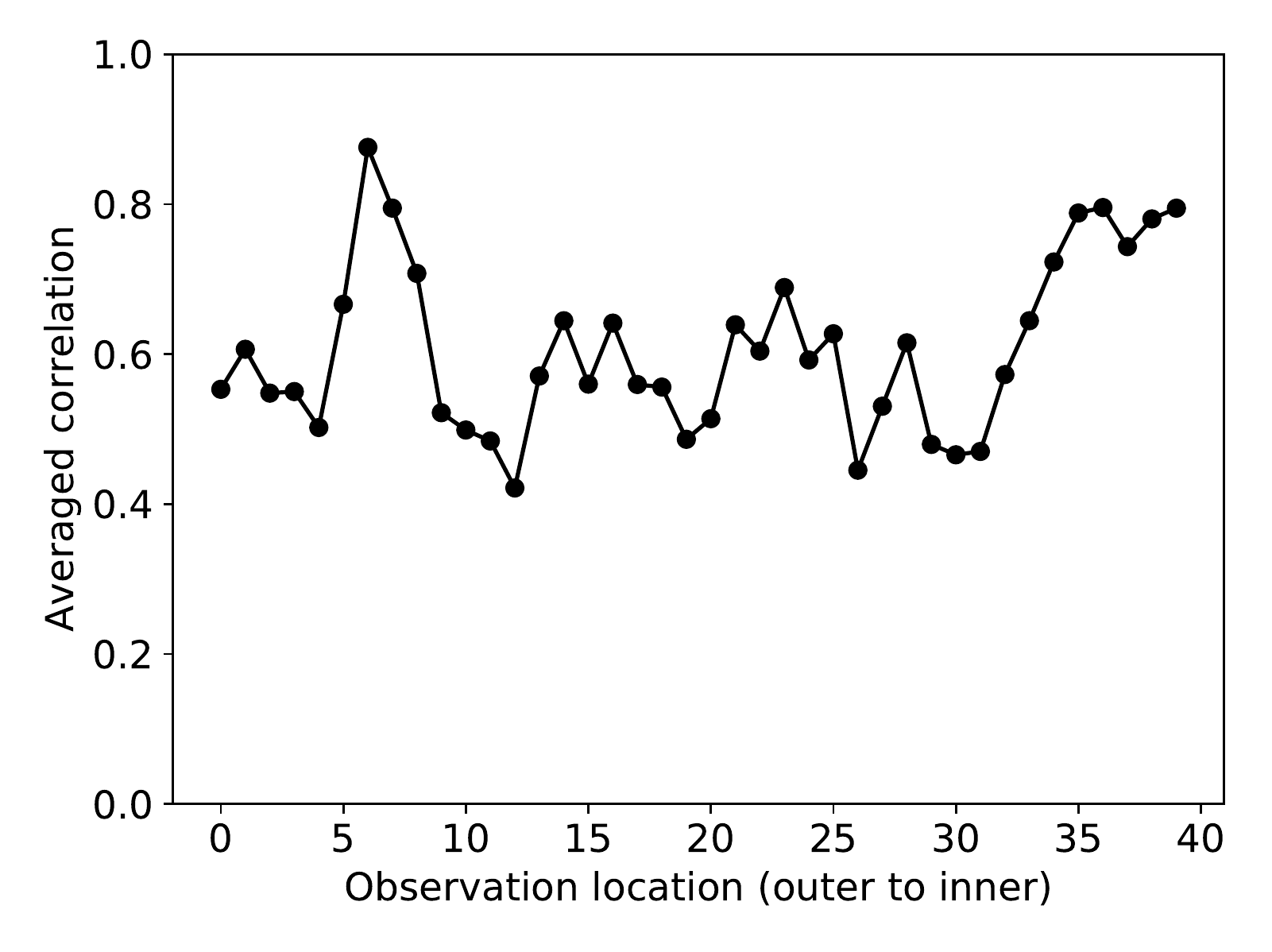}
	\caption{Averaged correlations between observed and the best synthetic light curves in each observation region. The horizontal axis indicates the observation location (from outside to inside).}
	\label{fig:corr}
\end{figure}
\begin{figure*}[tp]
	\centering
	\includegraphics[width=1\linewidth]{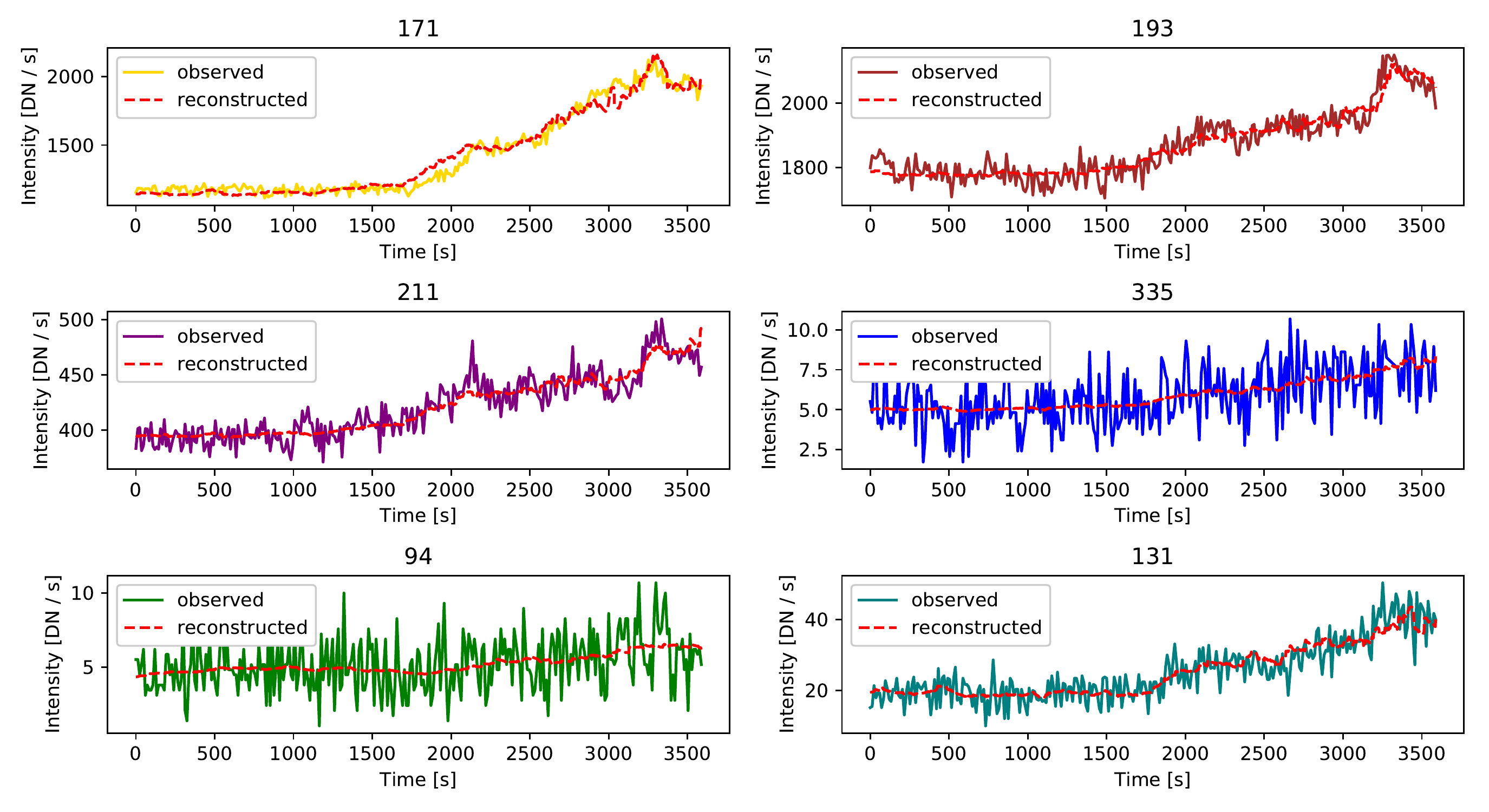}
	\caption{Sample of observed SDO/AIA light curves (solid lines) and synthetic ones which are optimized to reproduce the observation the most by GA (red dashed lines), which has the best average correlation ($\simeq 0.9$). }
	\label{fig:result_lc}
\end{figure*}
\begin{figure}[htp]
	\centering
	\includegraphics[width=1\linewidth]{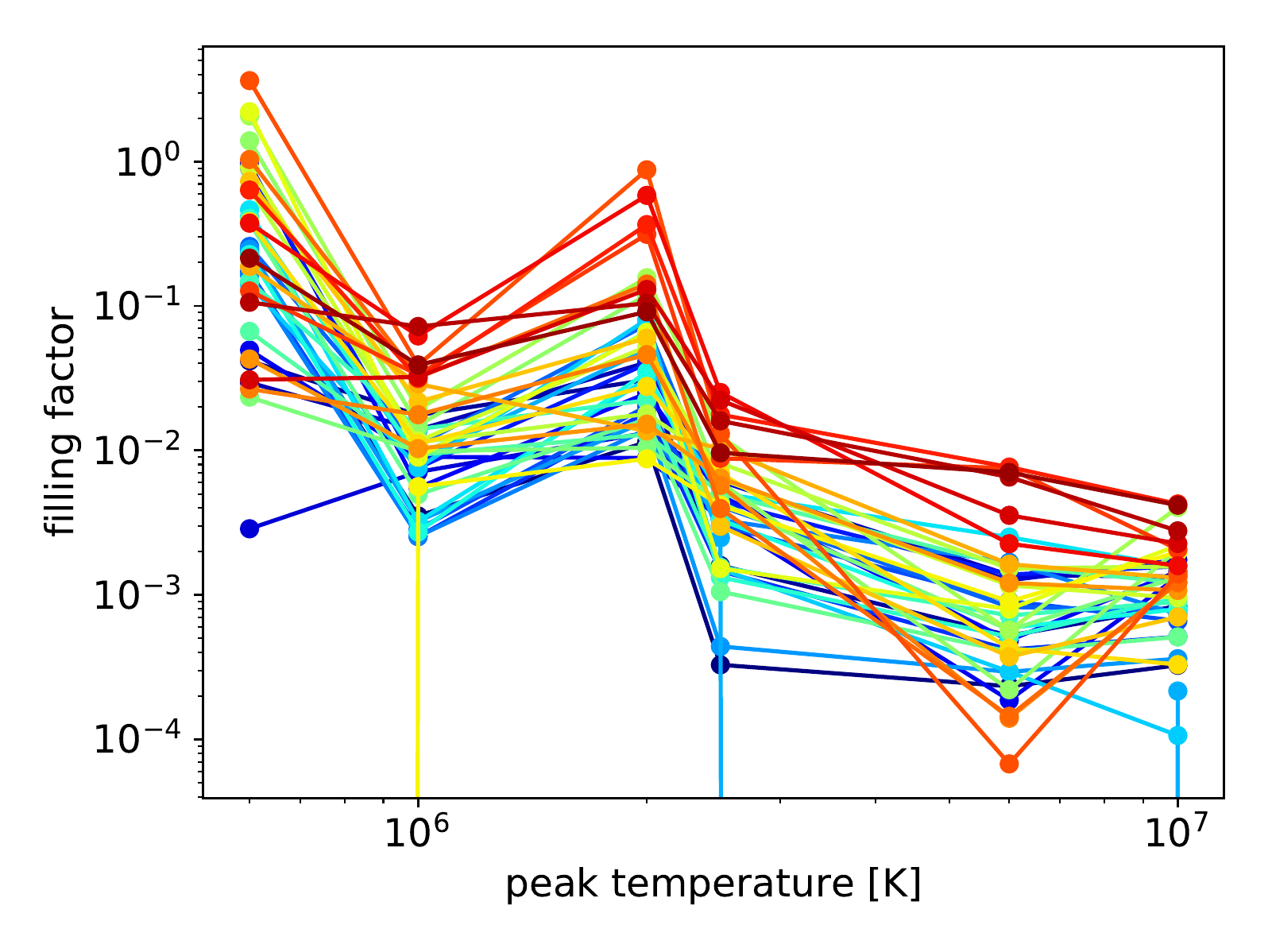}
	\caption{Filling factors of heated coronal loop in each observation location and filter. Horizontal axis represents the peak temperature of response function of each SDO/AIA channel. Colder (Warmer) color indicates farther distance from (closer to) the core.}
	\label{fig:fill}
\end{figure}

Figure~\ref{fig:corr} represents the average correlations between observed and optimized synthetic light curves. 
The horizontal axis indicates observation locations (red squares in figure~\ref{fig:AR}), from outside to inside. 
The correlations are approximately $0.4$ -- $0.9$, though they depend on the regions. 

Figure~\ref{fig:result_lc} shows an example of observed light curves obtained from SDO/AIA six channels (solid lines) and synthetic ones estimated by GA, which reproduces the observation best (dashed lines). 
These light curves are obtained from the region where the average correlation is the best ($\simeq 0.9$). 
Large fluctuations are reproduced accurately, however, smaller fluctuations other than photon noise are not reproduced well. 
\begin{figure*}[tp]
	\centering
	\includegraphics[width=1\linewidth]{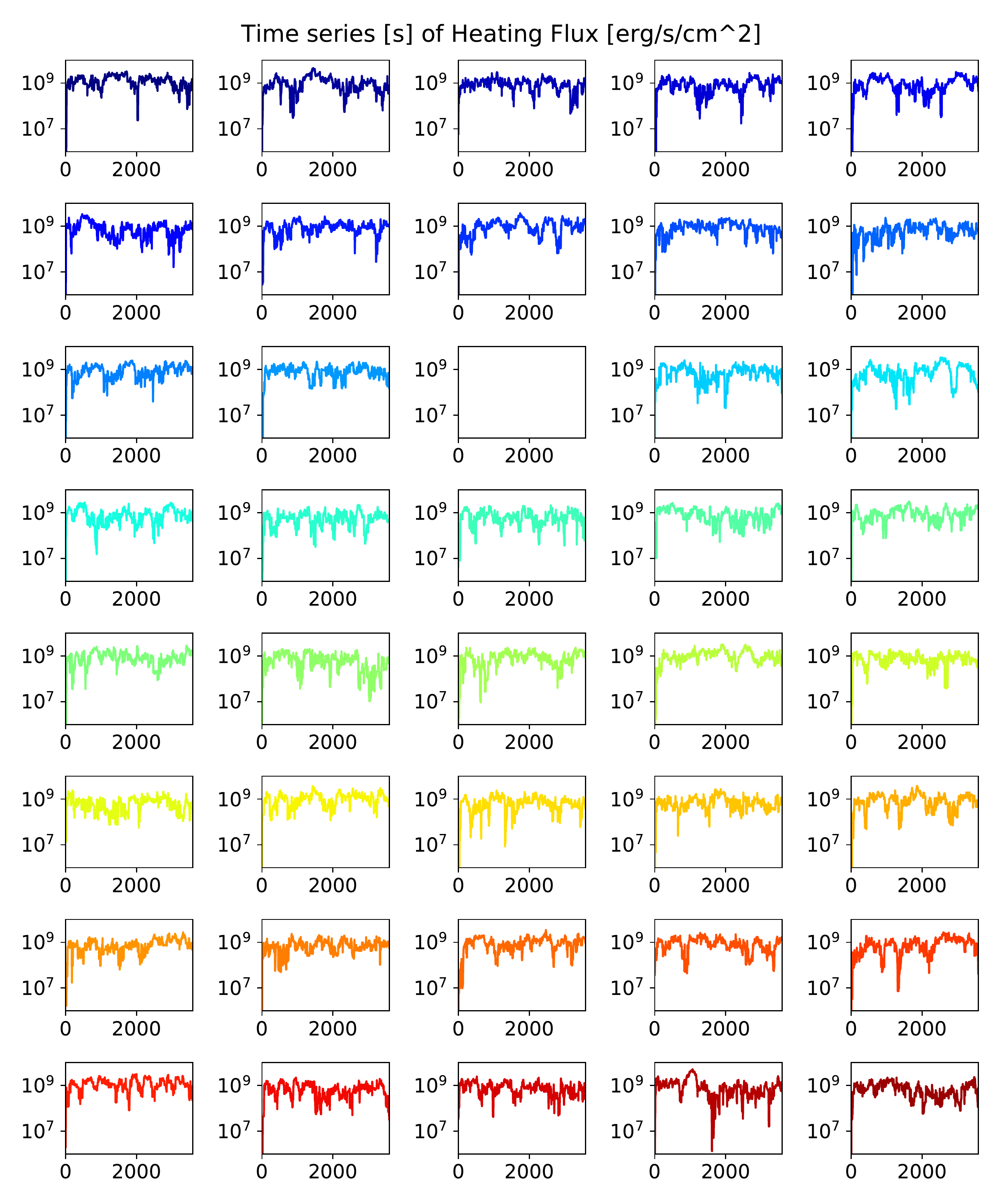}
	\caption{Time series of energy flux of flares in each observation region derived by GA. Colder (Warmer) color indicates farther distance from (closer to) the core. The panel without lines indicates that the GA determines that the calculations without flares from $t=2000$ to $t=5600$ reproduce the observed light curves best. }
	\label{fig:flux}
\end{figure*}

We define a cross sectional area along the loop of each observation region (red square in figure~\ref{fig:AR}) as $S_\mathrm{obs}=w^2$,
where $w$ represents the width of each red square. 
In this study, the size of each red square is $4 \times 4$ pixels, and hence, $w \simeq 1.75~\mathrm{Mm}$ and $S_\mathrm{obs} \simeq 3.1~\mathrm{Mm^2}$. 
We define the filling factor of the coronal loop $\phi$ as follows: 
\begin{equation}
\phi(\lambda)=\frac{l \times \Delta S_\mathrm{loop}(\lambda)}{S_\mathrm{obs}}
\end{equation}
where, $l$ represents the length of each gene. 
Figure~\ref{fig:fill} presents the estimated filling factors of all observation areas for each AIA filter. 
The warmer color indicates observation locations which are closer to the core. 
The horizontal axis represents the peak temperature of the response function of each AIA channel. 
As a result, the higher the plasma temperature, the lower the filling factor. 
In addition, the location closer to the core tends to have larger filling factor. 

The energy flux by flares can be described as:
\begin{equation}
	F(t) = \frac{1}{l}\sum_i^l H_{f}^{k_i}(t)~\mathrm{[erg/s/cm^2]}
\end{equation}
where $H_f^k$ is the flux caused by flares in simulation ID $k$. 
Each panel of figure~\ref{fig:flux} represents the estimated time series of energy flux by flares in each observation area. 
The color indicates the observation location as same as figure~\ref{fig:fill}. 
The panel without lines shows that the GA determines that the calculations without flares from $t=2000$ to $t=5600$ are the best to reproduce the observed light curves. 
Almost all observed regions are heated by flares intermittently, which have energy flux above the typical requirement \citep[$10^7~\mathrm{erg/s/cm^2}$:][]{Withbroe1977}. 

\begin{figure}[tp]
	\centering
	\includegraphics[width=1\linewidth]{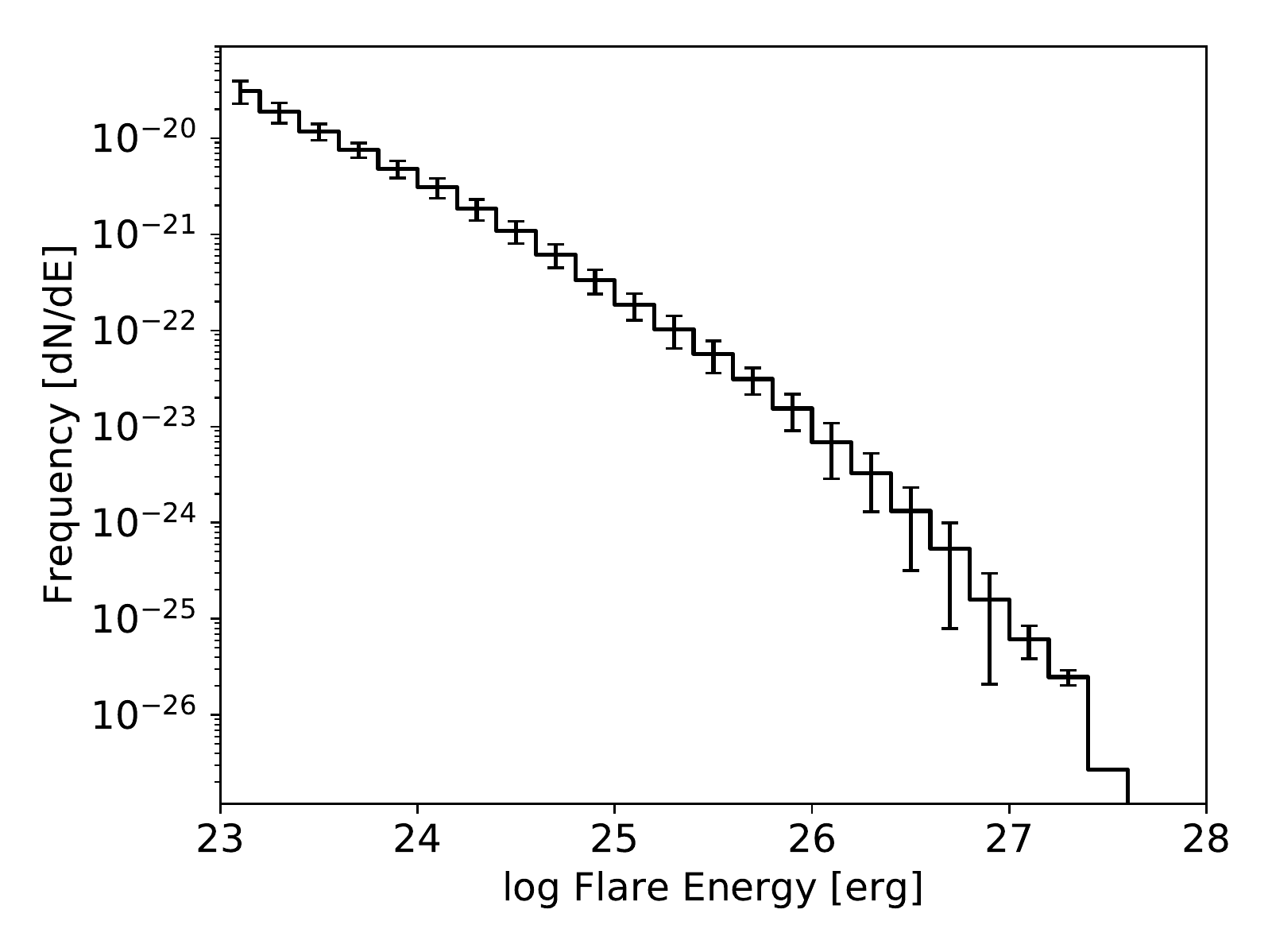}
	\caption{Frequency distribution of flares as a function of energy estimated by GA. vertical lines indicate errors which defined by the test described in the beginning of Section~\ref{sec:result}. The number of all detected flares is 37,458.}
	\label{fig:ene}
\end{figure}

\begin{figure}[tp]
	\centering
	\includegraphics[width=1\linewidth]{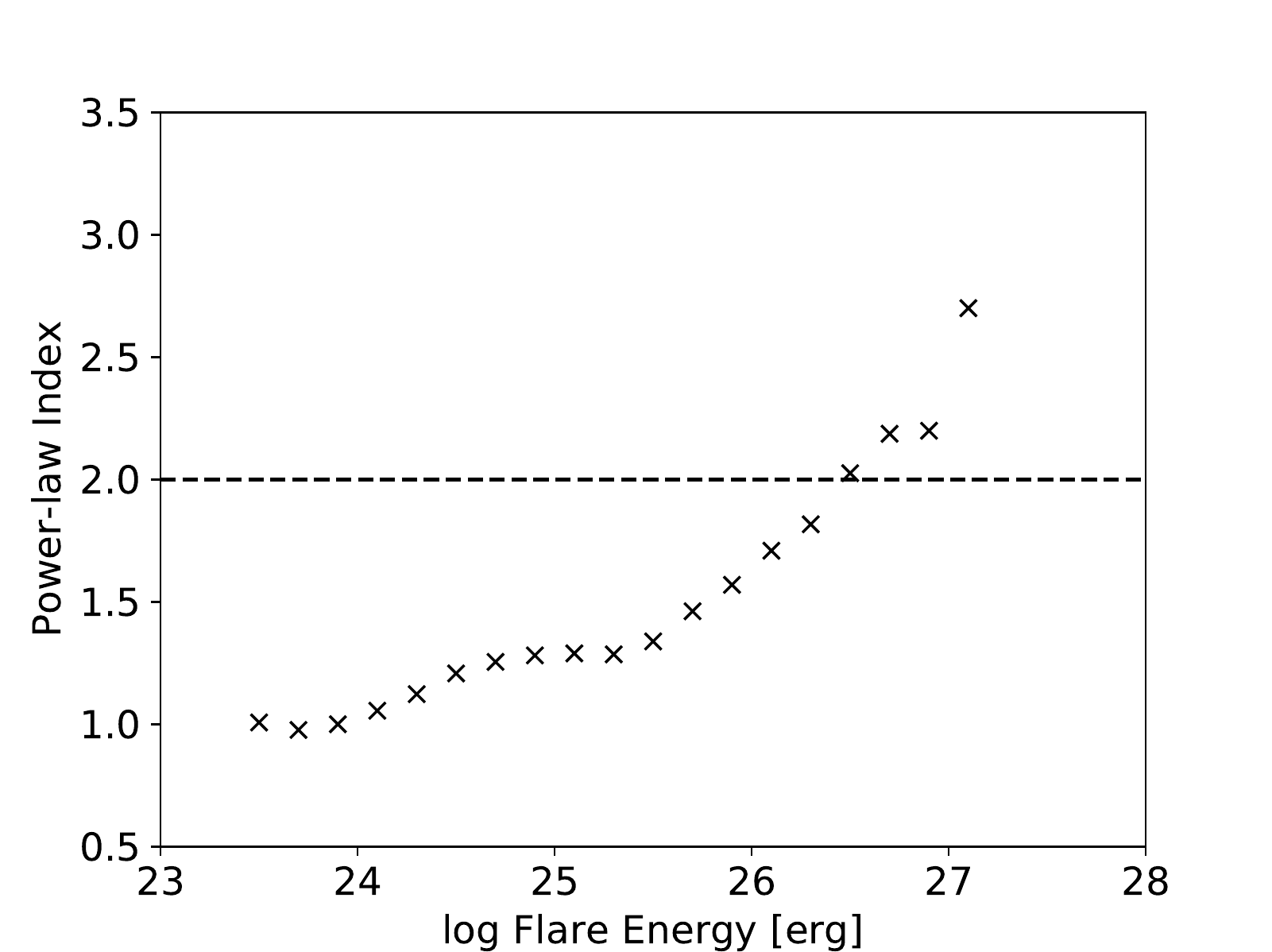}
	\caption{Series of power-law indices of the occurrence frequency distribution shown in Figure~\ref{fig:ene} in each energy bin. The size of each bin is the width that the flare energy on a logarithmic scale changes one. Dashed horizontal line indicates where power-law index becomes 2 which implies whether smaller flares are dominant in the coronal heating or not.}
	\label{fig:alphas}
\end{figure}

Figure~\ref{fig:ene} presents the frequency distribution of detected flares as a function of energy. 
Vertical lines indicate uncertainties defined by the test described in the beginning of this section as follow: 
\begin{equation}
	\sigma_f^\prime = N_\mathrm{obs} \times \Delta S_\mathrm{loop}(94) \times \sigma_f.
\end{equation}
where, $N_\mathrm{obs}$ represents the number of observation regions (40 in this study). 
The volume of the heated loop is necessary for the energy estimation; however, it depends on the filter. 
Therefore, in this study, we estimate the energy with $\Delta S_\mathrm{loop}(94)$, which is the most similar to the X-ray observational study of \citet{Shimizu1995}. 
The number of all detected flares is approximately $37,000$. 
Generally, the occurrence frequency distribution of flares as a function of energy show the power-law. 
However, the result does not have a single power-law index through the energy range. 
Therefore, we calculate the power-law index of the distribution in each energy bin as shown in Figure~\ref{fig:alphas}.
The size of each bin is defined as the width by which the flare energy on a logarithmic scale changes by one.
Horizontal axis indicates median of flare energy of each bin. 
In most cases, the power-law index increases with flare energy. 
The power-law index is greater than 2 in the energy range from $10^{26}$ to $10^{27}~\mathrm{erg}$ and above. 
On the other hand, the index is approximately 1 in the energy range smaller than $10^{25.5}~\mathrm{erg}$. 
This energy range is probably lower than the detection limit of our method because the index of simulation input is also approximately $1$. 

\begin{figure}[tp]
	\centering
	\includegraphics[width=1\linewidth]{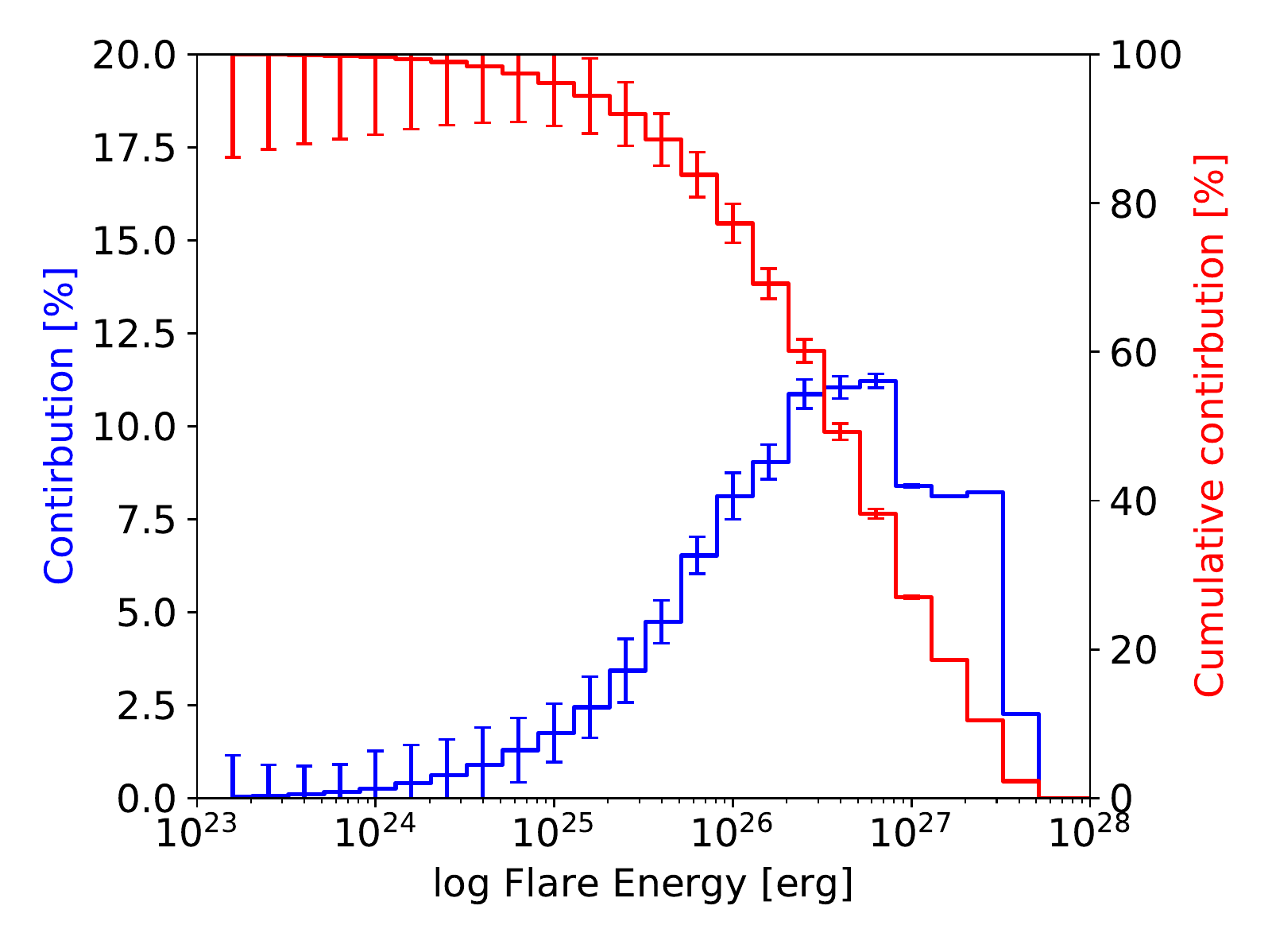}
	\caption{Contribution of flares to the coronal heating in each energy range (left axis, blue) and the cumulative contribution from high energy (right axis, red). Vertical lines indicate the errors calculated from the definition of the standard deviation $\sigma_f$ described in the beginning of Section~\ref{sec:result}.}
	\label{fig:cont}
\end{figure}

The blue line (left axis) in figure~\ref{fig:cont} represents the contribution of flares to heat the corona in each energy range. 
The red line (right axis) represents the cumulative contribution from high energy. 
Vertical lines indicate the errors calculated from $\sigma_f$ defined in the beginning of this section. 
Flares that contribute the most to the heating of the corona are those in the energy range of $10^{26} \lesssim E \lesssim 10^{27}~\mathrm{erg}$. 
Moreover, the $90\%$ of the coronal heating is done by flares that have energies greater than $10^{25}~\mathrm{erg}$.

\begin{figure}[tp]
	\centering
	\includegraphics[width=1\linewidth]{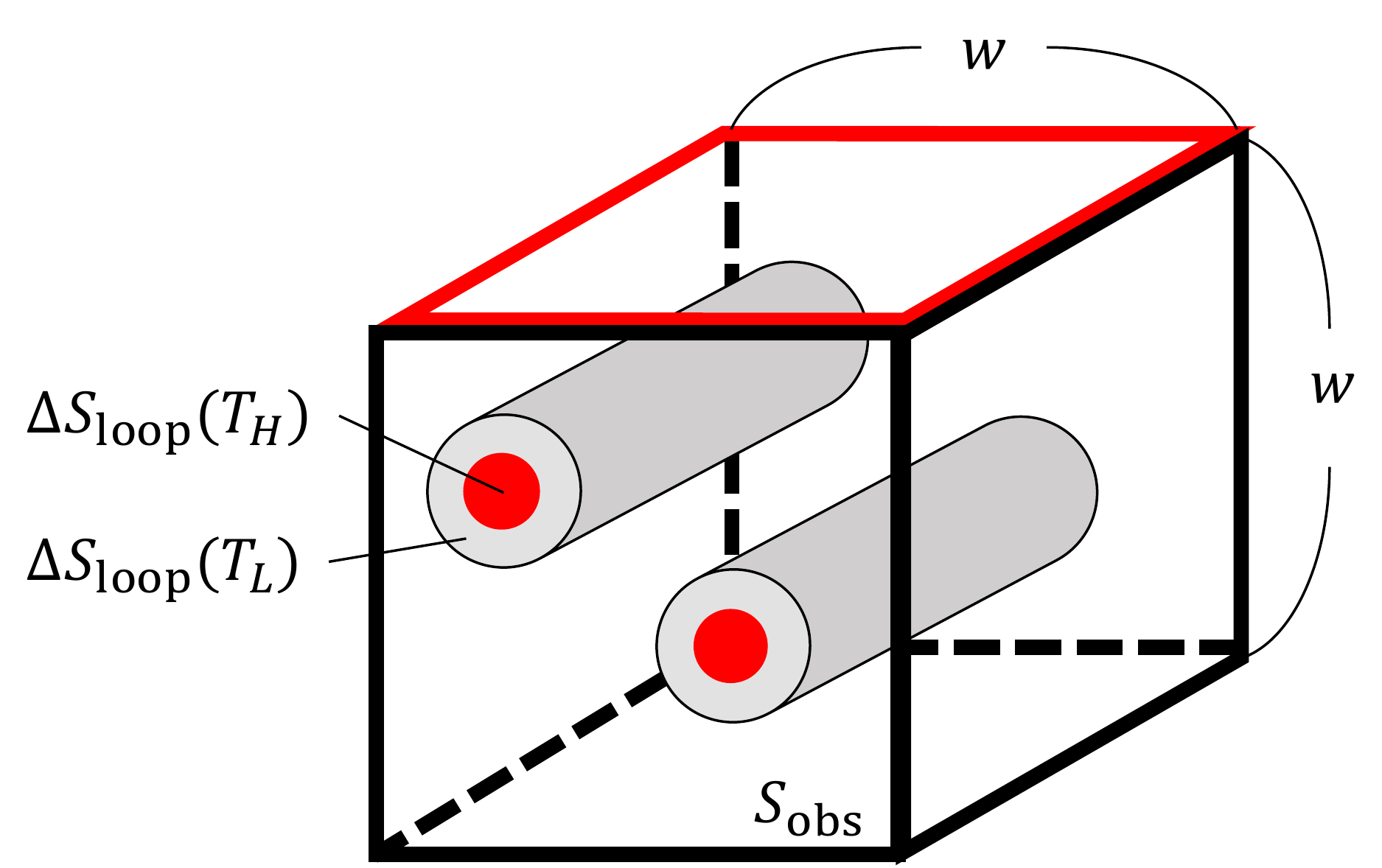}
	\caption{Schematic picture of coronal loops in the observation region. $\Delta S_\mathrm{loop}(T_H)$ and $\Delta S_\mathrm{loop}(T_L)$ represents the cross sectional areas of loops composed of hotter and cooler plasmas, respectively. Red square represents the observation region depicted in Figure~\ref{fig:AR}.}
	\label{fig:sc_fill}
\end{figure}

\section{Discussion and Summary} \label{sec:summary}
In this paper, we introduced a new method for detection and energy estimation of small-scale flares by using one-dimensional simulation and GA, which is a machine learning technique. 
We applied our method to the active-region coronal loop observation by SDO/AIA and obtained the time series of energy flux of flares, occurrence frequency distribution as a function of energy, and filling factors.  

The occurrence frequency distribution generally can be fitted by the power-law distribution, however, the power-law index of the derived distribution depends on the fitting range. 
As a result, the power-law index is found to be greater than 2 in flare energy range of $10^{26} < E < 10^{28}~\mathrm{erg}$ (Figure~\ref{fig:alphas}).
In this energy range, smaller flares are dominant in heating the corona, which is in line with the nanoflare heating model. 
On the other hand, the power-law index is approximately 1 in the energy range of smaller than $10^{25.5}~\mathrm{erg}$. 
This is probably caused by the detection limitation of our method because the distribution is similar to that of simulation inputs.
Moreover, we found that the coronal loops are heated by flares that have enough energy flux to heat the corona intermittently (Figure~\ref{fig:flux}). 
This is mainly because we calculate the fine structure of the loop that cannot be resolved. 
From figure~\ref{fig:fill}, we found that the volume that is heated by flares is much smaller than the observational resolution though it depends on the temperature and region. 
On the other hand, we found that the flares in energy range of $10^{26} \lesssim E \lesssim 10^{27}~\mathrm{erg}$ contribute to the coronal heating the most. 
In addition, $90\%$ of the energy flux comes from flares that release energies greater than $10^{25}~\mathrm{erg}$. 
It has been thought that smaller flares ($10^{23}~\textendash~10^{24}~\mathrm{erg}$) contribute the most to heat the corona because the energy flux of detected flares are not enough. 
There are some reasons why our result is incompatible with previous studies. 
We calculate the plasma evolution of coronal loops taking into account smaller loops than the observational resolution unlike previous studies. 
It is possible that the flare energy and occurrence frequency could be derived more accurately than previous studies by considering such a process.
This is because our energy estimation is based on physics unlike the method that defines a flare energy as a difference of thermal energy of a loop \citep[{\it e.g., \rm}][]{Shimizu1995}. 
However, our method has some shortcomings.
First, our results do not explain the balance of energy on the whole active region because we focus only on the coronal loop and not on the core region. 
Second, as we mentioned above, our method might be unable to detect small-scale flares which have an energy less than approximately $10^{24}~\mathrm{erg}$. 
Third, our combined synthetic light curves reproduce large fluctuations in the observation while smaller fluctuations are not reproduced well. 

We found that the filling factor increases as we get closer to the core, and the variation between the minimum and maximum is approximately three orders of magnitude. 
This is probably because the loops closer to the core can experience more magnetic pressure from the outer loops. 
This tendency is acceptable because the loops closer to the core are brighter than those in the outer region. 
As shown in figure~\ref{fig:fill}, the higher the peak temperature of each SDO/AIA filter, the lower the filling factor of coronal loops. 
This tendency is in agreement with the study of \citet{Sakamoto2009}, which compared the filling factors of the soft X-ray and EUV loops. 
This temperature dependence implies that the hotter coronal loops have a sparser structure than the cooler ones. 
Figure~\ref{fig:sc_fill} presents the schematic picture of coronal loops in each observation region. 
$\Delta S_\mathrm{loop}(T_H)$ (red circle) and $\Delta S_\mathrm{loop}(T_L)$ (gray circle) represents the cross-sectional areas of each loop composed of hotter and cooler plasmas, respectively. 
The red square corresponds to that depicted in Figure~\ref{fig:AR}.

The largest filling factor is greater than 100\%.
This can be caused by the assumption that the observational region is regarded as a cube. 
Coronal loops might be distributed along the line-of-sight direction deeper than observational width $w$.
On the other hand, in some cases, the filling factor is very small, approximately 0.01\%. 
Assuming $l$ of the elemental loops exist in each observation region, the half radius of each elemental loop $d$ can be described as: 
\begin{equation}
	d = \sqrt{\frac{S_\mathrm{obs} \phi}{\pi l}}.
\end{equation}
Therefore, in case of $0.01 \% < \phi < 100 \%$, 
\begin{equation}
	 10 \lesssim d \lesssim 990~\mathrm{[km]}.
\end{equation}
This result roughly corresponds to the maximum half radius of the elemental loop estimated by sounding rocket experiment, which is approximately $15~\mathrm{km}$ \citep{Peter2013}. 
$10~\mathrm{km}$ is roughly equivalent to $0.014~\mathrm{''}$ at the solar surface, which is about 2.3 \% of the resolution of SDO/AIA. 

To make large amounts of simulation results for GA, we use the one-dimensional coronal loop model which has a low computational cost. 
As mentioned in section~\ref{sec:intro}, the evolution of the coronal loop can be described in 1D when ignoring some processes. 
Strictly speaking, flares are magnetic reconnections between the loops that cannot be described using the one-dimensional model. 
Instead of this process, we directly substitute the flare heating term into the energy equation. 
However, making thousands of simulation results by using multidimensional model is unrealistic for current computational performance. 

We use CANS solar flare package to calculate a coronal loop heated by small-scale flares. 
Therefore we neglect some important processes in small-scale flare analysis such as non-equilibrium of ionization (NEI), radiative transfer, and non-thermal electron beams. 
\citet{Reale2008, Imada2011, Imada2015} suggest that NEI should have effects on the emission from the plasma which is impulsively heated by flares. 
Using the  HYDRAD model \citep[{\it e.g., \rm}][]{Bradshaw2010}, \citet{Bradshaw2011} studies a NEI effect on the coronal loop emissions for the case that a loop is heated by nanoflares. 
Because our simulation neglects NEI unlike HYDRAD, energy of small-scale flares may be estimated smaller than the actual and some small-scale events cannot be distinguished. 
Including this process to the simulation should make the analysis more accurate. 
\citet{Polito2018} suggest that the Si IV Doppler velocity at the chromosphere is a good indicator for detection of nanoflares ($\simeq 10^{24}~\mathrm{erg}$) from the analysis based on the RADYN model \citep[{\it e.g., \rm}][]{Allred2015}. 
It is necessary to validate the result by analyzing both coronal loops and chromospheres. 
\citet{Testa2014} also use the RADYN model and reveal that an observation of small and rapid variability of intensity and velocity at the loop foot point is consistent with heating by non-thermal electron beams generated by nanoflares. 
These processes should also be included in our analysis, however.

We must derive the combination of simulations which reproduce the observations the most. 
Therefore, we use GA, which is used to solve optimization and searching problems such as the traveling salesman problem. 
GA is suitable for statistical analysis, however, it has initial value dependence and may find only a local minimum. 
Moreover, it is difficult to prove that the combined synthetic light curve has only one combination because all combinations (${}_{5000}\mathrm{C}_{100} > 10^{200}$) must be calculated to prove that. 
In addition, GA has some free parameters such as length of each gene and the number of genes and iterations. 
The larger number of genes and iterations can make the correlation better, and thus, they should be as large as possible. 
On the other hand, the length of each gene defines the number of elemental loops in each observation region. 
In this paper, we assume that the number of loops in each macro pixel is 100 and try to estimate the best combination of simulations under this condition. 
However, the number of elemental loops in each identifiable loop is high and under discussion \citep[{\it e.g.,\rm}][]{Polito2018, Tajfirouze2016}. 
Therefore, it is necessary to analyze flares in various cases of the length of each gene. 

The authors thank K. Kusano for fruitful discussions. 
We would like to thank T. Yokoyama for his efforts in developing CANS package. 
This work was partially supported by the Grant-in-Aid for 17K14401 and 15H05816, and the Program for Leading Graduate Schools, ``PhD Professional: Gateway to Success in Frontier Asia'' by the Ministry of Education, Culture, Sports, Science and Technology. 
The {\it Solar Dynamics Observatory} is a part of NASA's Living with a Star program. 
A part of this study was carried out by using the computational resource of the Center for Integrated Data Science, Institute for Space-Earth Environmental Research, Nagoya University.


\bibliography{reference}{}

\begin{thebibliography}{}
\expandafter\ifx\csname natexlab\endcsname\relax\def\natexlab#1{#1}\fi
\providecommand{\url}[1]{\href{#1}{#1}}
\providecommand{\dodoi}[1]{doi:~\href{http://doi.org/#1}{\nolinkurl{#1}}}
\providecommand{\doeprint}[1]{\href{http://ascl.net/#1}{\nolinkurl{http://ascl.net/#1}}}
\providecommand{\doarXiv}[1]{\href{https://arxiv.org/abs/#1}{\nolinkurl{https://arxiv.org/abs/#1}}}

\bibitem[{{Allred} {et~al.}(2015){Allred}, {Kowalski}, \&
  {Carlsson}}]{Allred2015}
{Allred}, J.~C., {Kowalski}, A.~F., \& {Carlsson}, M. 2015, \apj, 809, 104,
  \dodoi{10.1088/0004-637X/809/1/104}

\bibitem[{{Antolin} \& {Shibata}(2010)}]{Antolin2010}
{Antolin}, P., \& {Shibata}, K. 2010, \apj, 712, 494,
  \dodoi{10.1088/0004-637X/712/1/494}

\bibitem[{{Aschwanden} {et~al.}(2000){Aschwanden}, {Tarbell}, {Nightingale},
  {Schrijver}, {Title}, {Kankelborg}, {Martens}, \& {Warren}}]{Aschwanden2000}
{Aschwanden}, M.~J., {Tarbell}, T.~D., {Nightingale}, R.~W., {et~al.} 2000,
  \apj, 535, 1047, \dodoi{10.1086/308867}

\bibitem[{{Benz} \& {Krucker}(2002)}]{Benz2002}
{Benz}, A.~O., \& {Krucker}, S. 2002, \apj, 568, 413, \dodoi{10.1086/338807}

\bibitem[{{Bradshaw} \& {Cargill}(2010)}]{Bradshaw2010}
{Bradshaw}, S.~J., \& {Cargill}, P.~J. 2010, \apj, 717, 163,
  \dodoi{10.1088/0004-637X/717/1/163}

\bibitem[{{Bradshaw} \& {Klimchuk}(2011)}]{Bradshaw2011}
{Bradshaw}, S.~J., \& {Klimchuk}, J.~A. 2011, \apjs, 194, 26,
  \dodoi{10.1088/0067-0049/194/2/26}

\bibitem[{{Cargill}(1994)}]{Cargill1994}
{Cargill}, P.~J. 1994, \apj, 422, 381, \dodoi{10.1086/173733}

\bibitem[{{Cargill} {et~al.}(2012){Cargill}, {Bradshaw}, \&
  {Klimchuk}}]{Cargill2012}
{Cargill}, P.~J., {Bradshaw}, S.~J., \& {Klimchuk}, J.~A. 2012, \apj, 752, 161,
  \dodoi{10.1088/0004-637X/752/2/161}

\bibitem[{{Christe} {et~al.}(2016){Christe}, {Glesener}, {Buitrago-Casas},
  {Ishikawa}, {Ramsey}, {Gubarev}, {Kilaru}, {Kolodziejczak}, {Watanabe},
  {Takahashi}, {Tajima}, {Turin}, {Shourt}, {Foster}, \&
  {Krucker}}]{Christe2016}
{Christe}, S., {Glesener}, L., {Buitrago-Casas}, C., {et~al.} 2016, Journal of
  Astronomical Instrumentation, 5, 1640005, \dodoi{10.1142/S2251171716400055}

\bibitem[{{Collura} {et~al.}(1988){Collura}, {Pasquini}, \&
  {Schmitt}}]{Collura1988}
{Collura}, A., {Pasquini}, L., \& {Schmitt}, J.~H.~M.~M. 1988, \aap, 205, 197

\bibitem[{{Datlowe} {et~al.}(1974){Datlowe}, {Elcan}, \&
  {Hudson}}]{Datlowe1974}
{Datlowe}, D.~W., {Elcan}, M.~J., \& {Hudson}, H.~S. 1974, \solphys, 39, 155,
  \dodoi{10.1007/BF00154978}

\bibitem[{{Dennis}(1985)}]{Dennis1985}
{Dennis}, B.~R. 1985, \solphys, 100, 465, \dodoi{10.1007/BF00158441}

\bibitem[{{Drake}(1971)}]{Drake1971}
{Drake}, J.~F. 1971, \solphys, 16, 152, \dodoi{10.1007/BF00154510}

\bibitem[{{Freeland} \& {Handy}(1998)}]{Freeland1998}
{Freeland}, S.~L., \& {Handy}, B.~N. 1998, \solphys, 182, 497,
  \dodoi{10.1023/A:1005038224881}

\bibitem[{{Hori} {et~al.}(1997){Hori}, {Yokoyama}, {Kosugi}, \&
  {Shibata}}]{Hori1997}
{Hori}, K., {Yokoyama}, T., {Kosugi}, T., \& {Shibata}, K. 1997, \apj, 489,
  426, \dodoi{10.1086/304754}

\bibitem[{{Hudson}(1991)}]{Hudson1991}
{Hudson}, H.~S. 1991, \solphys, 133, 357, \dodoi{10.1007/BF00149894}

\bibitem[{{Imada} {et~al.}(2015){Imada}, {Murakami}, \& {Watanabe}}]{Imada2015}
{Imada}, S., {Murakami}, I., \& {Watanabe}, T. 2015, Physics of Plasmas, 22,
  101206, \dodoi{10.1063/1.4932335}

\bibitem[{{Imada} {et~al.}(2011){Imada}, {Murakami}, {Watanabe}, {Hara}, \&
  {Shimizu}}]{Imada2011}
{Imada}, S., {Murakami}, I., {Watanabe}, T., {Hara}, H., \& {Shimizu}, T. 2011,
  \apj, 742, 70, \dodoi{10.1088/0004-637X/742/2/70}

\bibitem[{{Imada} \& {Zweibel}(2012)}]{Imada2012}
{Imada}, S., \& {Zweibel}, E.~G. 2012, \apj, 755, 93,
  \dodoi{10.1088/0004-637X/755/2/93}

\bibitem[{{Ishikawa} {et~al.}(2017){Ishikawa}, {Glesener}, {Krucker},
  {Christe}, {Buitrago-Casas}, {Narukage}, \& {Vievering}}]{Ishikawa2017}
{Ishikawa}, S., {Glesener}, L., {Krucker}, S., {et~al.} 2017, Nature Astronomy,
  1, 771, \dodoi{10.1038/s41550-017-0269-z}

\bibitem[{{Jess} {et~al.}(2019){Jess}, {Dillon}, {Kirk}, {Reale},
  {Mathioudakis}, {Grant}, {Christian}, {Keys}, {Krishna Prasad}, \&
  {Houston}}]{Jess2019}
{Jess}, D.~B., {Dillon}, C.~J., {Kirk}, M.~S., {et~al.} 2019, \apj, 871, 133,
  \dodoi{10.3847/1538-4357/aaf8ae}

\bibitem[{{Kawai} {et~al.}(2020){Kawai}, {Imada}, {Nishimoto}, {Watanabe}, \&
  {Kawate}}]{Kawai2020}
{Kawai}, T., {Imada}, S., {Nishimoto}, S., {Watanabe}, K., \& {Kawate}, T.
  2020, Journal of Atmospheric and Solar-Terrestrial Physics, 205, 105302,
  \dodoi{10.1016/j.jastp.2020.105302}

\bibitem[{{Klimchuk}(2006)}]{Klimchuk2006}
{Klimchuk}, J.~A. 2006, \solphys, 234, 41, \dodoi{10.1007/s11207-006-0055-z}

\bibitem[{{Klimchuk}(2015)}]{Klimchuk2015}
---. 2015, Philosophical Transactions of the Royal Society of London Series A,
  373, 20140256, \dodoi{10.1098/rsta.2014.0256}

\bibitem[{{Klimchuk} {et~al.}(2008){Klimchuk}, {Patsourakos}, \&
  {Cargill}}]{Klimchuk2008}
{Klimchuk}, J.~A., {Patsourakos}, S., \& {Cargill}, P.~J. 2008, \apj, 682,
  1351, \dodoi{10.1086/589426}

\bibitem[{{Kobayashi} {et~al.}(2014){Kobayashi}, {Cirtain}, {Winebarger},
  {Korreck}, {Golub}, {Walsh}, {De Pontieu}, {DeForest}, {Title}, {Kuzin},
  {Savage}, {Beabout}, {Beabout}, {Podgorski}, {Caldwell}, {McCracken},
  {Ordway}, {Bergner}, {Gates}, {McKillop}, {Cheimets}, {Platt}, {Mitchell}, \&
  {Windt}}]{Kobayashi2014}
{Kobayashi}, K., {Cirtain}, J., {Winebarger}, A.~R., {et~al.} 2014, \solphys,
  289, 4393, \dodoi{10.1007/s11207-014-0544-4}

\bibitem[{{Lemen} {et~al.}(2012){Lemen}, {Title}, {Akin}, {Boerner}, {Chou},
  {Drake}, {Duncan}, {Edwards}, {Friedlaender}, \& {Heyman}}]{Lemen2012}
{Lemen}, J.~R., {Title}, A.~M., {Akin}, D.~J., {et~al.} 2012, \solphys, 275,
  17, \dodoi{10.1007/s11207-011-9776-8}

\bibitem[{{Lin} {et~al.}(1984){Lin}, {Schwartz}, {Kane}, {Pelling}, \&
  {Hurley}}]{Lin1984}
{Lin}, R.~P., {Schwartz}, R.~A., {Kane}, S.~R., {Pelling}, R.~M., \& {Hurley},
  K.~C. 1984, \apj, 283, 421, \dodoi{10.1086/162321}

\bibitem[{{Ogawara} {et~al.}(1991){Ogawara}, {Takano}, {Kato}, {Kosugi},
  {Tsuneta}, {Watanabe}, {Kondo}, \& {Uchida}}]{Ogawara1991}
{Ogawara}, Y., {Takano}, T., {Kato}, T., {et~al.} 1991, \solphys, 136, 1,
  \dodoi{10.1007/BF00151692}

\bibitem[{{Parker}(1983)}]{Parker1983}
{Parker}, E.~N. 1983, \apj, 264, 635, \dodoi{10.1086/160636}

\bibitem[{{Parker}(1988)}]{Parker1988}
---. 1988, \apj, 330, 474, \dodoi{10.1086/166485}

\bibitem[{{Parnell} \& {Jupp}(2000)}]{Parnell2000}
{Parnell}, C.~E., \& {Jupp}, P.~E. 2000, \apj, 529, 554, \dodoi{10.1086/308271}

\bibitem[{{Pesnell} {et~al.}(2012){Pesnell}, {Thompson}, \&
  {Chamberlin}}]{Pesnell2012}
{Pesnell}, W.~D., {Thompson}, B.~J., \& {Chamberlin}, P.~C. 2012, \solphys,
  275, 3, \dodoi{10.1007/s11207-011-9841-3}

\bibitem[{{Peter} {et~al.}(2013){Peter}, {Bingert}, {Klimchuk}, {de Forest},
  {Cirtain}, {Golub}, {Winebarger}, {Kobayashi}, \& {Korreck}}]{Peter2013}
{Peter}, H., {Bingert}, S., {Klimchuk}, J.~A., {et~al.} 2013, \aap, 556, A104,
  \dodoi{10.1051/0004-6361/201321826}

\bibitem[{{Polito} {et~al.}(2018){Polito}, {Testa}, {Allred}, {De Pontieu},
  {Carlsson}, {Pereira}, {Go{\v{s}}i{\'c}}, \& {Reale}}]{Polito2018}
{Polito}, V., {Testa}, P., {Allred}, J., {et~al.} 2018, \apj, 856, 178,
  \dodoi{10.3847/1538-4357/aab49e}

\bibitem[{{Priest}(1978)}]{Priest1978}
{Priest}, E.~R. 1978, \solphys, 58, 57, \dodoi{10.1007/BF00152555}

\bibitem[{{Reale}(2014)}]{Reale2014}
{Reale}, F. 2014, Living Reviews in Solar Physics, 11, 4,
  \dodoi{10.12942/lrsp-2014-4}

\bibitem[{{Reale} \& {Orlando}(2008)}]{Reale2008}
{Reale}, F., \& {Orlando}, S. 2008, \apj, 684, 715, \dodoi{10.1086/590338}

\bibitem[{{Rosner} {et~al.}(1978){Rosner}, {Tucker}, \& {Vaiana}}]{Rosner1978}
{Rosner}, R., {Tucker}, W.~H., \& {Vaiana}, G.~S. 1978, \apj, 220, 643,
  \dodoi{10.1086/155949}

\bibitem[{{Sakamoto} {et~al.}(2009){Sakamoto}, {Tsuneta}, \&
  {Vekstein}}]{Sakamoto2009}
{Sakamoto}, Y., {Tsuneta}, S., \& {Vekstein}, G. 2009, \apj, 703, 2118,
  \dodoi{10.1088/0004-637X/703/2/2118}

\bibitem[{{Schmelz} {et~al.}(2001){Schmelz}, {Scopes}, {Cirtain}, {Winter}, \&
  {Allen}}]{Schmelz2001}
{Schmelz}, J.~T., {Scopes}, R.~T., {Cirtain}, J.~W., {Winter}, H.~D., \&
  {Allen}, J.~D. 2001, \apj, 556, 896, \dodoi{10.1086/321588}

\bibitem[{{Shakhovskaya}(1989)}]{Shakhovskaya1989}
{Shakhovskaya}, N.~I. 1989, \solphys, 121, 375, \dodoi{10.1007/BF00161707}

\bibitem[{{Shimizu}(1995)}]{Shimizu1995}
{Shimizu}, T. 1995, \pasj, 47, 251

\bibitem[{{Shimizu} {et~al.}(1994){Shimizu}, {Tsuneta}, {Acton}, {Lemen},
  {Ogawara}, \& {Uchida}}]{Shimizu1994}
{Shimizu}, T., {Tsuneta}, S., {Acton}, L.~W., {et~al.} 1994, \apj, 422, 906,
  \dodoi{10.1086/173782}

\bibitem[{{Shimizu} {et~al.}(1992){Shimizu}, {Tsuneta}, {Acton}, {Lemen}, \&
  {Uchida}}]{Shimizu1992}
{Shimizu}, T., {Tsuneta}, S., {Acton}, L.~W., {Lemen}, J.~R., \& {Uchida}, Y.
  1992, \pasj, 44, L147

\bibitem[{{Snodgrass} \& {Ulrich}(1990)}]{Snodgrass1990}
{Snodgrass}, H.~B., \& {Ulrich}, R.~K. 1990, \apj, 351, 309,
  \dodoi{10.1086/168467}

\bibitem[{{Spitzer}(1956)}]{Spitzer1956}
{Spitzer}, L. 1956, {Physics of Fully Ionized Gases}

\bibitem[{{Tajfirouze} {et~al.}(2016){Tajfirouze}, {Reale}, {Petralia}, \&
  {Testa}}]{Tajfirouze2016}
{Tajfirouze}, E., {Reale}, F., {Petralia}, A., \& {Testa}, P. 2016, \apj, 816,
  12, \dodoi{10.3847/0004-637X/816/1/12}

\bibitem[{{Testa} {et~al.}(2013){Testa}, {De Pontieu}, {Mart{\'\i}nez-Sykora},
  {DeLuca}, {Hansteen}, {Cirtain}, {Winebarger}, {Golub}, {Kobayashi}, \&
  {Korreck}}]{Testa2013}
{Testa}, P., {De Pontieu}, B., {Mart{\'\i}nez-Sykora}, J., {et~al.} 2013, \apj,
  770, L1, \dodoi{10.1088/2041-8205/770/1/L1}

\bibitem[{{Testa} {et~al.}(2014){Testa}, {De Pontieu}, {Allred}, {Carlsson},
  {Reale}, {Daw}, {Hansteen}, {Martinez-Sykora}, {Liu}, {DeLuca}, {Golub},
  {McKillop}, {Reeves}, {Saar}, {Tian}, {Lemen}, {Title}, {Boerner},
  {Hurlburt}, {Tarbell}, {Wuelser}, {Kleint}, {Kankelborg}, \&
  {Jaeggli}}]{Testa2014}
{Testa}, P., {De Pontieu}, B., {Allred}, J., {et~al.} 2014, Science, 346,
  1255724, \dodoi{10.1126/science.1255724}

\bibitem[{{Tiwari} {et~al.}(2019){Tiwari}, {Panesar}, {Moore}, {De Pontieu},
  {Winebarger}, {Golub}, {Savage}, {Rachmeler}, {Kobayashi}, {Testa}, {Warren},
  {Brooks}, {Cirtain}, {McKenzie}, {Morton}, {Peter}, \& {Walsh}}]{Tiwari2019}
{Tiwari}, S.~K., {Panesar}, N.~K., {Moore}, R.~L., {et~al.} 2019, \apj, 887,
  56, \dodoi{10.3847/1538-4357/ab54c1}

\bibitem[{{Tsuneta} {et~al.}(1991){Tsuneta}, {Acton}, {Bruner}, {Lemen},
  {Brown}, {Caravalho}, {Catura}, {Freeland}, {Jurcevich}, \&
  {Morrison}}]{Tsuneta1991}
{Tsuneta}, S., {Acton}, L., {Bruner}, M., {et~al.} 1991, \solphys, 136, 37,
  \dodoi{10.1007/BF00151694}

\bibitem[{{Vekstein} \& {Katsukawa}(2000)}]{Vekstein2000}
{Vekstein}, G., \& {Katsukawa}, Y. 2000, \apj, 541, 1096,
  \dodoi{10.1086/309480}

\bibitem[{{Vesecky} {et~al.}(1979){Vesecky}, {Antiochos}, \&
  {Underwood}}]{Vesecky1979}
{Vesecky}, J.~F., {Antiochos}, S.~K., \& {Underwood}, J.~H. 1979, \apj, 233,
  987, \dodoi{10.1086/157462}

\bibitem[{{Warren} {et~al.}(2008){Warren}, {Ugarte-Urra}, {Doschek}, {Brooks},
  \& {Williams}}]{Warren2008}
{Warren}, H.~P., {Ugarte-Urra}, I., {Doschek}, G.~A., {Brooks}, D.~H., \&
  {Williams}, D.~R. 2008, \apjl, 686, L131, \dodoi{10.1086/592960}

\bibitem[{{Withbroe} \& {Noyes}(1977)}]{Withbroe1977}
{Withbroe}, G.~L., \& {Noyes}, R.~W. 1977, \araa, 15, 363,
  \dodoi{10.1146/annurev.aa.15.090177.002051}

\end{thebibliography}
\bibliographystyle{aasjournal}



\end{document}